\documentclass[fleqn,10pt]{wlscirep}
\usepackage[utf8]{inputenc}
\usepackage[T1]{fontenc}
\usepackage{lineno}
\usepackage{xcolor}

\title{Experienced partisan segregation across patterns of mobility behavior in US cities}

\author[1,2,*]{Marco Tonin}
\author[1]{Michele Tizzoni}
\author[2]{Bruno Lepri}
\author[3]{Esteban Moro}

\affil[1]{Department of Sociology and Social Research, University of Trento, Via Verdi 26, Trento, I-38122, Italy}
\affil[2]{Mobile and Social Computing Lab (MobS), Fondazione Bruno Kessler (FBK), Via Sommarive 18, Trento, I-38123, Italy}
\affil[3]{Network Science Institute, Department of Physics, Northeastern University, Boston, 02115, MA, USA}

\affil[*]{marco.tonin@unitn.it}

\begin{abstract}
Residential and social sorting along political lines structure Americans’ daily interactions, often limiting encounters with opposing viewpoints.
Here, we combine high-resolution anonymized mobility data from 11 major U.S. metropolitan areas with precinct-level voting results to quantify and examine individuals’ experienced partisan segregation in urban areas.
We find that residential context provides a strong baseline for political exposure, but experienced partisan segregation is also shaped by everyday mobility behavior, since individuals living in similar residential environments can experience different levels of partisan segregation depending on their daily routines.
In particular, individuals who travel farther encounter more politically diverse environments, especially in highly segregated areas, indicating that mobility range moderates the relationship between residential and experienced segregation, while exploring a wider range of places does not reduce partisan segregation and may even reinforce it.
Moreover, place segregation largely reflects political geography, yet categories of points of interest have distinct segregation patterns, with community places and social businesses tending to attract more left-leaning visitors, reflecting visitation patterns rather than local accessibility alone.
Our findings show that experienced partisan segregation extends beyond residential geography and is also associated with the mobility behavior and visitation patterns that structure people’s daily lives.
\end{abstract}
\begin{document}
% \linenumbers
\flushbottom
\maketitle

\vspace{-1cm}
\vspace{0.5cm}
%\thispagestyle{empty}

\section*{Introduction}
In the United States, social interactions in both physical and digital spaces often reflect political sorting, manifested in neighborhoods and geographic space~\cite{brown_measurement_2021,kaplan_partisan_2022,brown_sources_2025,brown_relationship_2025}, daily interactions and contacts~\cite{gentzkow_ideological_2011,tonin_physical_2025}, workplaces~\cite{chinoy2024political,frake_political_2026}, activity spaces~\cite{zhang_human_2023}, and social media platforms\cite{cinelli_echo_2021,di_martino_ideological_2025}.
Such divisions can limit exposure to differing opinions and perspectives, shape the spread of information, reinforce affective polarization, and influence political behavior, as social networks and contexts in which individuals are embedded are associated with social and political participation\cite{bond_61-million-person_2012,campbell_social_2013}, as well as the reinforcement or alignment of political opinions and voting choices\cite{lazer_coevolution_2010,sinclair_social_2012,strother_college_2021,tonin_physical_2025,brown_partisan_2025}.
Within these contexts, contact can take different forms and intensities~\cite{nathan_context_2023}, ranging from brief and casual encounters, which can also influence political attitudes and behavior~\cite{enos_causal_2014,enos_space_2017}, to more sustained and selective interactions, as described in the literature on contact theory\cite{allport1954nature}.

The political divide in space is often associated with residential patterns~\cite{brown_measurement_2021,kaplan_partisan_2022}, with Democrats tending to cluster in metropolitan areas, while Republicans are more segregated in rural areas.
These patterns are further reinforced by generational change in urban contexts and ideological conformity in rural areas~\cite{brown_sources_2025}, but not by political preferences for relocation, even if individuals prefer to live in more politically homogeneous communities~\cite{gimpel_seeking_2015,mummolo_why_2017}.
As a result, individuals are often embedded in politically homogeneous residential environments, with limited exposure to those with opposing political views.

However, where people live only partially determines the environments they encounter in daily life.
Individual mobility, defined as where people go, how far they travel, and which places they visit, can shape exposure to diverse groups and contexts and is associated with social outcomes such as segregation and inequalities~\cite{wang_urban_2018,moro_mobility_2021,athey_estimating_2021,luca_crime_2023,nilforoshan_human_2023,xu_using_2025,liao_socio-spatial_2025,tonin_physical_2025}, public health~\cite{aleta_modelling_2020,lucchini_living_2021}, job opportunities, and economic outcomes~\cite{wang_infrequent_2024}. 
These mobility patterns also reflect lifestyle choices, as Democrats and Republicans can differ systematically in their routines, time-use, place preferences, and social relationships~\cite{dellaposta_why_2015,chen_effect_2018,bertrand_coming_2023,talaifar_lifestyle_2025}.
At the same time, mobility behavior is also shaped by external shocks~\cite{aleta_modelling_2020,lucchini_living_2021,yabe_behavioral_2023} and life-course events~\cite{centellegher_job_2025}.
Thus, mobility and behavioral patterns may influence experienced partisan segregation, even among individuals living in similar residential contexts.

In addition to differences in mobility behavior, not all activity spaces contribute equally to experienced partisan exposure.
Social infrastructures, defined as places where people are more likely to interact and build trust and social ties, are associated with social capital\cite{putnam2000bowling} in urban settings, reducing inequality and enhancing participation and civic life~\cite{klinenberg2018palaces,latham_social_2019,fraser_trust_2022}. 
Although these locations should theoretically foster cross-partisan exposure and interaction, greater accessibility to social infrastructures is, in practice, positively associated with votes for Democratic candidates and higher voter turnout\cite{fraser_trust_2022}.
Yet, accessibility does not always translate into use, and observed patterns of partisan exposure may reflect differences in mobility and visitation behavior rather than differences in spatial availability alone.

Here, we examine political segregation across places and the built environment, with a specific focus on social infrastructures, and how individual mobility behavior relates to experienced partisan exposure in urban areas.
Specifically, we first compute activity space and individual-level segregation and examine variation across 11 metropolitan areas, namely Boston, Chicago, Dallas, Detroit, Los Angeles, Miami, New York, Philadelphia, San Francisco, Seattle, and Washington, DC.
Second, we assess whether mobility behavior moderates or mediates the relationship between residential and experienced segregation, asking whether and how patterns of daily movement amplify or mitigate exposure to political differences.
Third, we examine the relationship between experienced partisan segregation and visitation patterns in social infrastructures as central places where people build social ties and capital.

To this end, we combine high-resolution anonymized large-scale GPS mobility data with precinct-level voting results from the 2016 U.S. presidential election. 
Our dataset covers 11 major U.S. metropolitan areas, enabling fine-grained measurement of individual exposure to partisan environments across both residential and activity spaces. 
For each individual, we measure residential and experienced political segregation and characterize mobility behavior using metrics such as place exploration, which is the tendency to visit new places, and the average distance traveled. 
This approach allows us to analyze how daily mobility shapes exposure to politically diverse environments.

Our findings show that the residential context does not fully account for the political environments individuals experience in their daily lives. 
Instead, individual mobility behavior and visitation patterns also shape experienced partisan segregation across activity spaces.
While the residential context is a major constraint, individual mobility behavior in terms of travel range moderates, rather than mediates, residential partisan segregation. 
Individuals who travel farther are exposed to more politically mixed encounters, even when they live in segregated areas.
However, when controlling for distance, place exploration and the tendency to visit new places do not reduce partisan segregation and may even reinforce it.
Additionally, experienced partisan segregation varies across types of activity spaces.
Place segregation strongly reflects the geography of the vote, and categories of points of interest exhibit distinct patterns of segregation that slightly decrease with increasing catchment range.
In particular, community places and social businesses, such as libraries, museums, bars, and other locations that foster social interaction and social capital, tend to be visited more by left-leaning individuals in most of the cities considered.
Our results indicate that this pattern is primarily associated with differences in visitation behavior rather than with accessibility or local availability alone.
Overall, these findings advance a behavioral understanding of experienced partisan segregation and disentangle how mobility behavior can shape exposure to political differences beyond the residential context and across activity spaces.

\section*{Results}
Using high-resolution anonymized mobility data collected between October 2016 and April 2017, we quantify experienced partisan segregation in urban contexts and investigate its relationship with mobility behavior.
We first examine the political composition of visitors to each point of interest (POI).
Second, we measure each user’s experienced segregation based on the political composition of the POIs they visit and assess how mobility behavior relates to cross-partisan exposure. 
Finally, we focus on social infrastructures, places that foster interaction and social capital, and examine the visitation patterns associated with experienced partisan segregation in these places. 

\begin{figure*}[t!]
\centering
\includegraphics[width=1\textwidth]{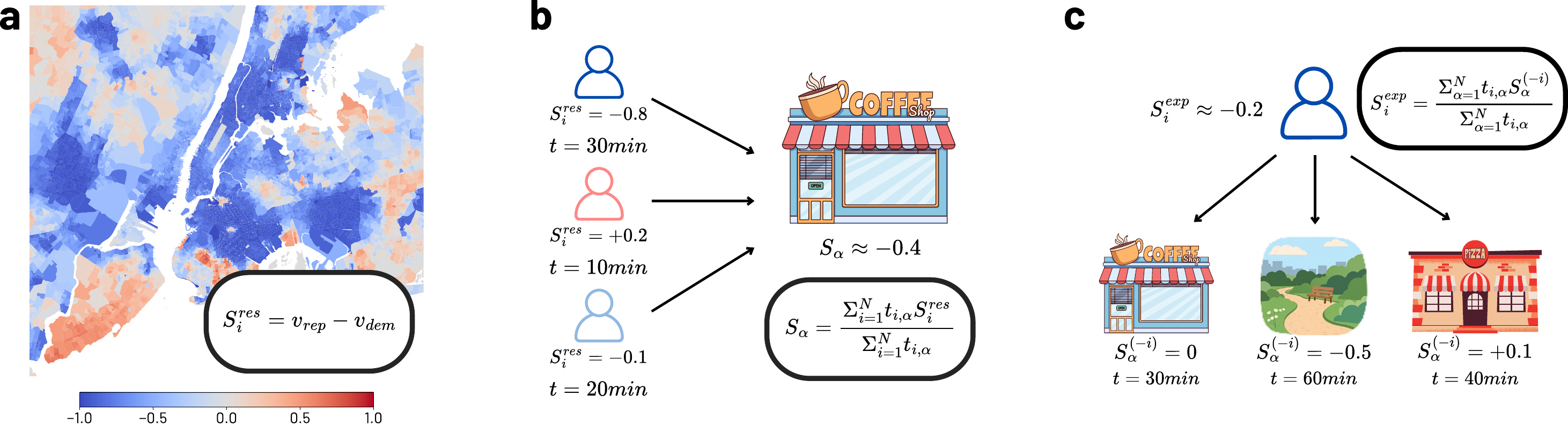} 
\caption{\textbf{Definitions of residential, place, and experienced partisan segregation.} \textbf{(a)} Individual residential segregation is computed based on the political vote in the precinct where the individual lives, taking the difference between shares of votes of republicans and democrats at the 2016 presidential election. Values range from -1 (Democratic segregation, blue) and +1 (Republican segregation, red). Part of the New York metropolitan area is represented in the map; \textbf{(b)} Place segregation is computed based on the time-weighted composition of the visitors, considering their residential partisan segregation. Here, stylized individuals are colored based on their residential segregation, $S_{i}^{res}$; \textbf{(c)} Individual experienced segregation is computed as the time-weighted place segregation of the locations they visit, using a leave-out measure to not consider self-exposure.}
\label{fig:1}
\end{figure*}

We integrate mobility data across 11 U.S. Core-Based Statistical Areas (CBSAs) with precinct-level results from the 2016 U.S. presidential election to infer individuals' residential political exposure based on their place of residence (see Materials and Methods). 
We define residential exposure as the difference between the Republican and Democratic vote shares (Fig.~\ref{fig:1}a).
This measure accounts for third-party votes since the combined Republican and Democratic share can be less than $1$. 
Residential political exposure ranges from $-1$, corresponding to a precinct where all voters support Democrats, to $1$, corresponding to a precinct where all voters support Republicans.

We then combine mobility and political data to measure both place- and individual-level experienced segregation in urban contexts. 
For each point of interest, $\alpha$, we compute its political composition as the time-weighted residential exposure of its visitors, $S_\alpha$ (Fig.~\ref{fig:1}b).
Similarly, we quantify the experienced partisan segregation of each individual, $S_i^{\text{exp}}$, as the time-weighted segregation of the POIs they visit using a leave-one-out measure that exclude the individuals' own contribution to the place segregation of each visited location (Fig.~\ref{fig:1}c). 
Thus, each individual is characterized by their residential political exposure and their experienced partisan segregation.
This approach provides a comprehensive framework to capture partisan mixing and segregation across millions of visits in diverse urban environments.

\subsection*{Place segregation reflects political geography}
\begin{figure*}[t!]
\centering
\includegraphics[width=1\textwidth]{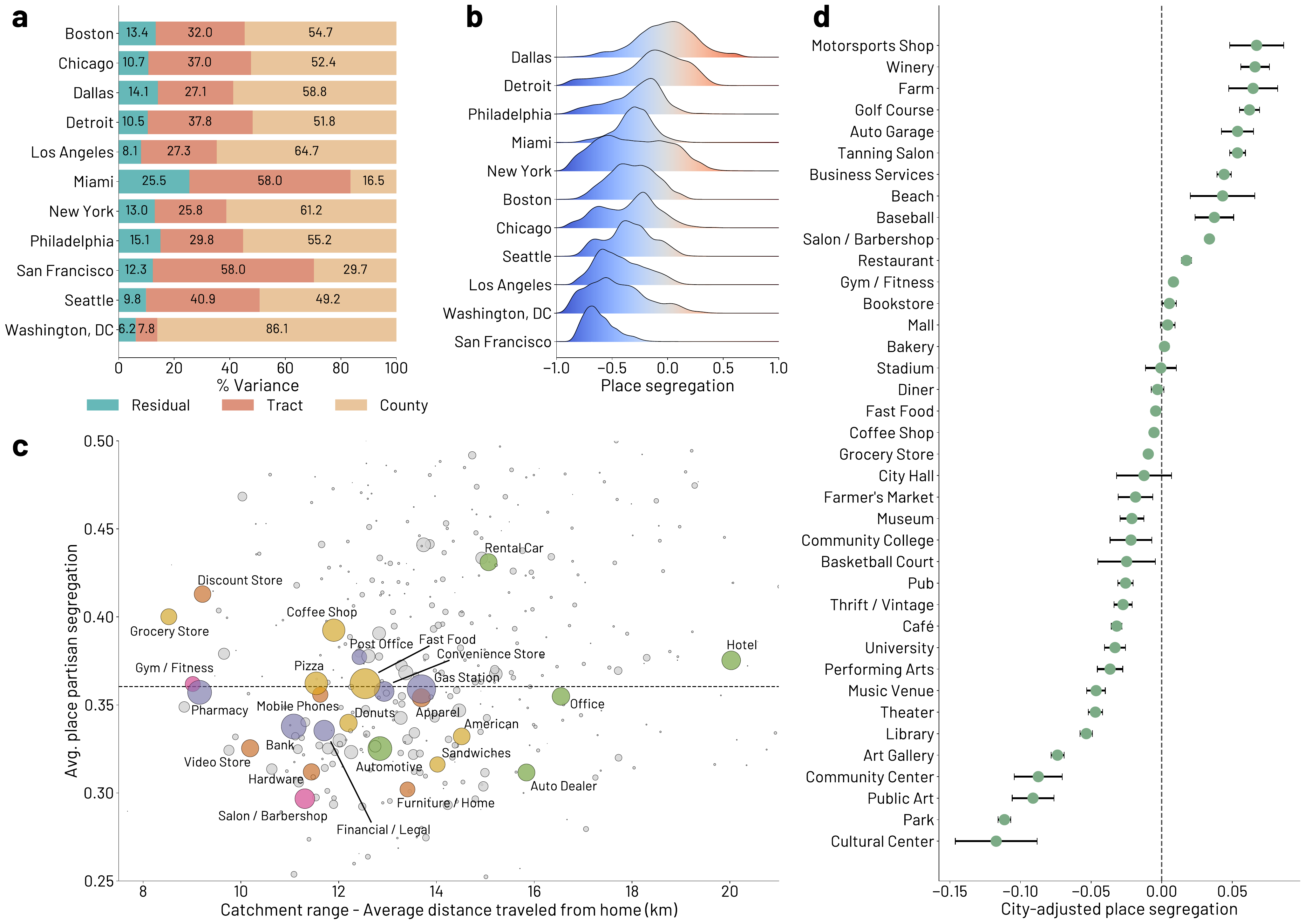} 
\caption{\textbf{Patterns of place segregation across metropolitan areas.} \textbf{(a)} Variance Partition Coefficients (VPCs) from the multilevel regression models show the modest share of within-areas variation in place segregation; \textbf{(b)} Place segregation distribution across the 11 U.S. metropolitan areas considered. Values range from -1, when a place is visited only by left-leaning individuals, to 1, when a place is visited only by right-leaning individuals; \textbf{(c)} Average place segregation by category slightly decreases across catchment range, defined as the average distance traveled to reach an activity space ($r = -0.16$). Here, we consider the absolute value of place segregation, with values between 0 and 1. The dashed horizontal line represents the average place segregation across categories; \textbf{(d)} Considering widely spread social places, some categories (with negative values) are visited more by left-leaning individuals, while others (with positive values) are visited more by right-leaning voters. Mean values are adjusted by subtracting the median place segregation in the city where the point of interest is located. Error bars represent the standard errors.}
\label{fig:2}
\end{figure*}

The location of activity spaces strongly reflects the political composition of their visitors. 
Political place segregation is highly correlated with the precinct-level voting outcomes ($r = 0.81$), suggesting that the political composition of places reflects the electoral context.
Moreover, places within the same Census Tract tend to share similar levels of segregation, showing that activity spaces are not randomly mixed and have homogeneous partisan compositions at the tract level.
We quantify the variance in place segregation that is not explained by differences between geographic areas using multilevel regression analyses with random intercepts with two nested geographic levels, County and Census Tract.
In most cities, the residual variance computed through the Variance Partition Coefficients (VPCs) from the multilevel regression models ranges from $6.2\%$ in Washington, DC, to $15.1\%$ in Philadelphia, with the exception of Miami, where it reaches $25.5\%$ (Fig.~\ref{fig:2}a), while differences between counties and tracts within counties account for the remaining variance. 

Besides this general pattern, the range of place-based political segregation varies by metropolitan area (Fig.~\ref{fig:2}b). 
Reflecting election results, places in most of the cities under study are visited by Democrats, with distributions skewed toward Democratic segregation. 
Some metropolitan areas, such as Dallas and Detroit, show greater mixing of partisan groups across activity spaces, with an average $S_\alpha$ equal to $-0.01$ and $-0.12$ respectively.
Finally, in cities such as Dallas, Detroit, and New York, the distributions are more widely spread, suggesting that activity spaces differ more strongly in their partisan composition and that residents can potentially encounter more diverse political environments based on their mobility behavior.
% Overall, these patterns mirror the underlying political geography of each metropolitan area.

The absolute value of political segregation of places, with values between 0 (full mixing) and 1 (full segregation), varies across categories of activity spaces (Fig.~\ref{fig:2}c) and slightly decreases with the catchment range (Pearson's correlation, $r = -0.16)$, defined as the average distance between individuals' home locations and the points of interest they visit.
While daily necessities such as grocery stores, discount stores, and pharmacies tend to have high average partisan segregation and short catchment ranges, categories with a medium catchment range (about 12 kilometers from home) show a wide spread in segregation levels.
For example, fast food restaurants, coffee shops, and post offices display higher segregation compared to salons, barber shops, or automotive services, despite similar distances from home. 
At the upper end of the catchment range, categories such as hotels and offices tend to show average segregation levels.
From this perspective, the figure shows how both distance and activity type shape the partisan composition of places people visit in daily life.

Interestingly, when considering the directionality of political segregation, some categories are more frequently visited by Democrats, while others are visited more by Republicans.
Fig.~\ref{fig:2}d shows place segregation after accounting for each city's expected segregation value by subtracting the median place segregation within the corresponding city. 
Negative values represent left-leaning categories, while positive values represent right-leaning ones. 
The selected categories include popular activity spaces for daily necessities and leisure. 
Categories such as parks ($-0.12$), community centers ($-0.11$), libraries ($-0.07$), and museums ($-0.03$) tend to attract more left-leaning visitors, while farms ($+0.08$), wineries ($+0.07$), baseball courts ($+0.02$), and golf courses ($+0.06$) are more commonly visited by right-leaning individuals.
In general, these findings highlight category-specific differences in political exposure shaped by different types of locations.

Overall, these patterns show that the political segregation of places reflects the underlying political geography and that everyday environments can potentially shape opportunities for political exposure and cross-group interaction.

\subsection*{Travel range, rather than exploration, reduces experienced partisan segregation}
People living in the same area can experience different levels of political segregation in their daily lives. Although contextual factors such as county and census tract account for a substantial share of the variation, they do not fully determine individuals' exposure to political groups. This suggests that, even if residential segregation constrains opportunities for political exposure, individual mobility behavior also contributes to shaping experienced partisan segregation.

%People living in the same area can experience different levels of political segregation in their daily lives. Although contextual factors such as county and census tract account for a substantial share of the variation, they do not fully determine individuals' exposure to political groups. This suggests that, even if residential segregation constrains opportunities for exposure, individual behavior and mobility patterns contribute to shaping partisan exposure. 

To quantify these dynamics, we estimate multilevel regression models with random intercepts nested at two geographic levels, county and census tract, and decompose the variance in experienced segregation into contextual and individual components. Consistent with the patterns of place partisan segregation, county and tract differences account for most of the variance in experienced segregation, with within-area variance ranging from $8.6\%$ in Washington, DC to $26\%$ in Miami (SI, Fig. S1). This remaining within-area variation suggests that individuals living in similar residential environments may nevertheless experience different political environments through their daily mobility.

%To quantify these dynamics, we estimate multilevel regression models with random intercepts nested at two geographic levels, county and census tract, and decompose the variance in experienced segregation into contextual and individual components. Consistent with the patterns of place partisan segregation, our findings show that county and tract differences account for most of the variance in experienced segregation, with within-areas variance ranging from $9.2\%$ in Washington, DC to $30.3\%$ in Miami (SI, Fig. S1a).

\begin{figure*}[t!]
\centering
\includegraphics[width=1\textwidth]{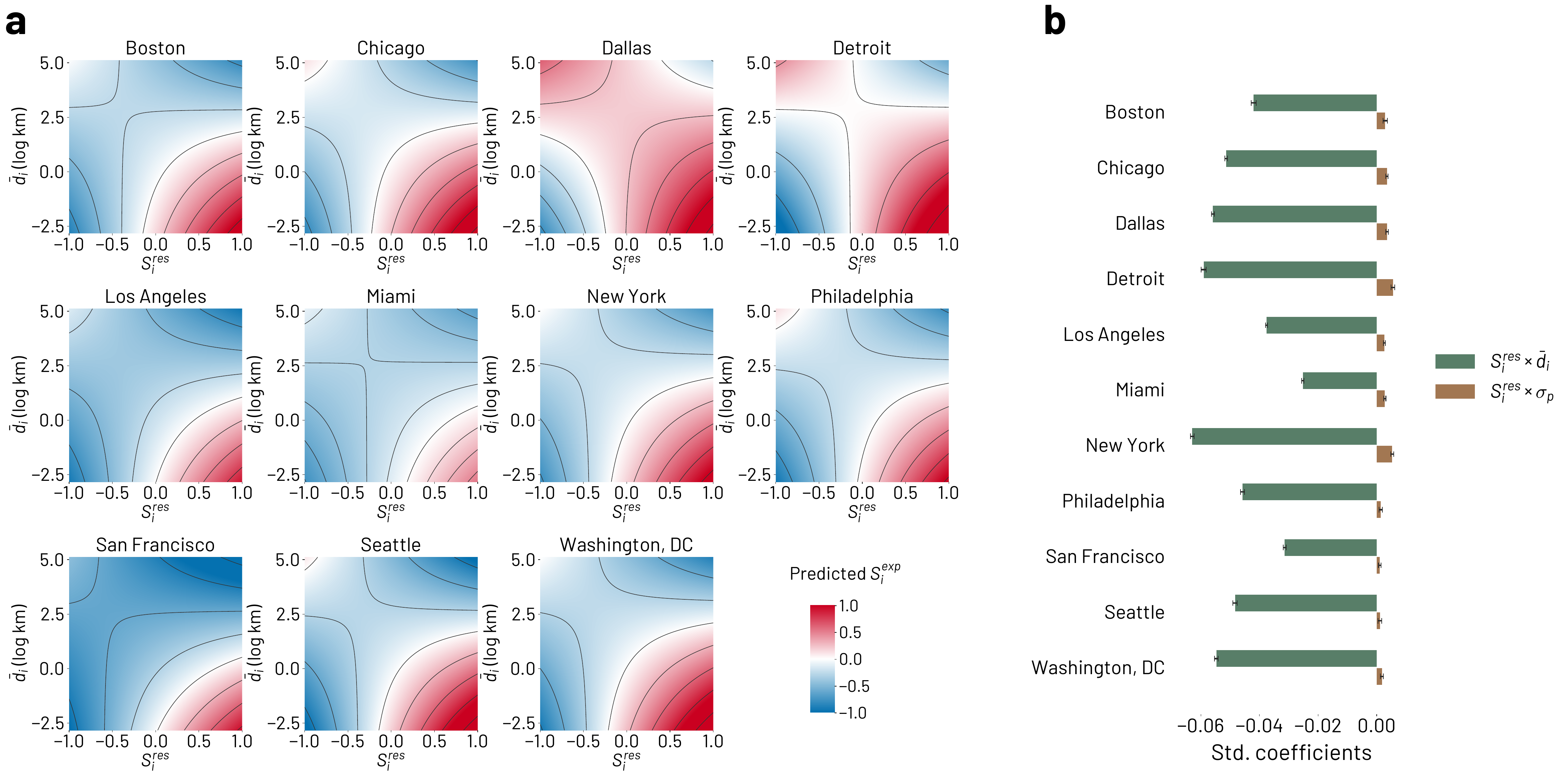} 
\caption{\textbf{Mobility behavior moderates residential partisan segregation.} \textbf{(a)} Moderation effects of the interaction term between residential partisan segregation and distance traveled on experienced segregation across CBSAs. The average distance traveled, $\bar{d}_i$, is computed as the logarithm of the mean distance in kilometers. Patterns are consistent across metropolitan areas with different predicted $S_{i}^{exp}$ values, reflecting differences political geography; \textbf{(b)} Standardized coefficients of the interaction terms are consistent across metropolitan areas. Distance traveled ($\bar{d}_i$, green) moderates the relationship between residential and experienced partisan segregation, whereas place exploration ($\sigma_p$, brown) does not.}
\label{fig:3}
\end{figure*}

Mobility behavior accounts for part of this within-area heterogeneity. We characterize mobility using two complementary measures: place exploration, $\sigma_p$, defined as the tendency to visit new places, and the time-weighted average distance from home, $\bar{d}_i$, which measures the average distance traveled to reach activity spaces. We also control for socioeconomic characteristics, including ethnicity, educational attainment, and income. To understand the role of mobility, we distinguish between two possible mechanisms (SI, Fig. S2). Mobility may act as a \emph{mediator}, explaining how residential context translates into experienced segregation, or as a \emph{moderator}, changing the strength of the relationship between residential and experienced segregation. Details of the analyses are provided in the Materials and Methods and SI, Section S3.

%Mobility patterns account for part of this within-area heterogeneity. To understand how human behavior and mobility shape experienced segregation, we model the interplay between experienced segregation, residential sorting, and mobility patterns. We characterize mobility behavior using two different metrics: place exploration, $\sigma_p$, defined as the tendency to visit new places, and the time-weighted average distance from home, $\bar{d}_i$, which measures the average distance traveled to reach an activity space. We also control for socioeconomic characteristics, including ethnicity, educational attainment, and income.

Not all forms of mobility contribute equally to reducing experienced partisan segregation. Figure~\ref{fig:3}a shows the predicted experienced segregation as a function of residential segregation and average distance traveled. The contour plots indicate that greater mobility range attenuates the impact of residential segregation on experienced segregation, particularly for individuals living in highly segregated residential environments. Across all 11 CBSAs, increasing travel distance is consistently associated with a weaker relationship between residential and experienced partisan segregation, with standardized interaction coefficients ranging from $-0.025$ (Miami) to $-0.063$ (New York) (Fig.~\ref{fig:3}b and SI, Table S5).

%First, we test two distinct mechanisms to understand whether mobility behavior acts as a mediator or a moderator (see path diagrams in the SI, Fig. S2). In the mediation framework, mobility patterns are treated as mediators of the relationship between residential and experienced political segregation. The hypothesis is that where people live shapes mobility patterns, both in terms of how much they travel and the diversity of the places they visit, which subsequently influences experienced segregation. Mediation allows us to disentangle the direct and indirect effects of residential segregation. The direct path quantifies the effects of residential segregation on experienced segregation, while the indirect paths include the effects of residential segregation on mobility patterns and, in turn, the influence of mobility patterns on experienced segregation (SI, Fig. S2a). Second, we test mobility patterns as moderators. In this case, mobility patterns can either amplify or reduce the effect of residential segregation on experienced segregation (SI, Fig. S2b). Details about model specification can be found in the Materials and Methods section and in the Supplementary Information.

By contrast, greater exploration does not translate into greater political mixing. The corresponding analysis using place exploration is reported in SI Fig. S3. After controlling for travel distance, the interaction between residential segregation and place exploration is positive and statistically significant, suggesting that exploratory mobility does not reduce, and may even reinforce, experienced partisan segregation. This contrasts with previous findings on experienced income segregation, where individuals with more exploratory mobility tended to experience less segregation~\cite{moro_mobility_2021,yabe_behavioral_2023}. Together, these findings indicate that, for political mixing, where people travel matters more than how many different places they visit.

The mediation analysis supports this interpretation. Across cities, the indirect pathways are small and close to zero (SI, Table S2), indicating that mobility contributes only marginally to the translation of residential segregation into experienced segregation. Instead, most of the association between residential and experienced segregation is direct. Mobility primarily changes the strength of this relationship: travel distance moderates the effect of residential segregation on experienced segregation, whereas place exploration does not. Including both interaction terms improves model performance across metropolitan areas (SI, Table S3). Regression summaries are reported in SI Tables S4 and S5.

Overall, these findings show that not all forms of mobility are equally associated with political mixing. Individuals who travel farther from home are exposed to more politically diverse environments, especially when they live in highly segregated neighborhoods. Unlike experienced income segregation, however, simply exploring a wider variety of places does not reduce experienced partisan segregation and may even reinforce it. Together, these results suggest that the spatial reach of everyday mobility, rather than exploratory behavior itself, is the primary behavioral mechanism through which individuals can partially escape residential political segregation.

\subsection*{Patterns of political mixing and segregation in social infrastructures}

\begin{figure*}[t!]
\centering
\includegraphics[width=1\textwidth]{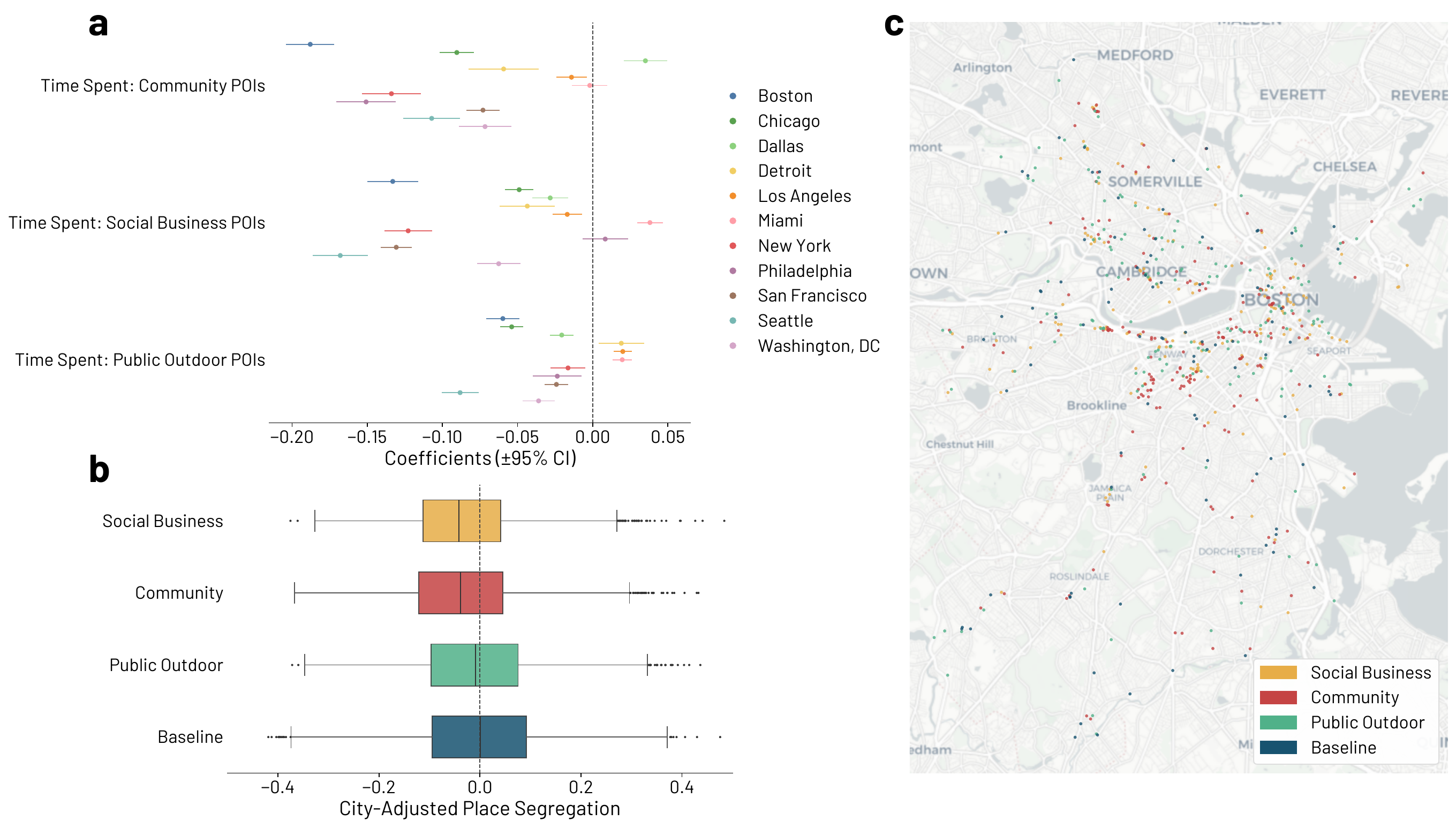} 
\caption{\textbf{Partisan segregation in social infrastructure locations.} \textbf{(a)} Overall, spending time in community and social business places is associated with greater exposure to left-leaning individuals, with the exception of Dallas and Miami. Negative values of the coefficients are associated with greater exposure to left-leaning Democrats, while positive values with right-leaning individuals; \textbf{(b)} Social businesses and community places, on average, tend to be more visited by left-leaning individuals. Place segregation is adjusted by accounting for the median value in each city; \textbf{(c)} Distribution of social infrastructures and baseline places (i.e., grocery stores, supermarkets, and post offices) in the cities of Boston, Cambridge, and Somerville.}
\label{fig:4}
\end{figure*}

Political mixing and sorting in social infrastructures, places where people are more likely to encounter, interact, and build social ties, vary across cities. 
To better understand these patterns, we distinguish between structural opportunity and accessibility, captured by the local availability of social infrastructures, and behavioral patterns, reflected in the places individuals choose to visit.
To this end, we estimate the relationship between experienced partisan segregation and visit patterns, controlling for local accessibility.

Drawing on the definitions and taxonomy in the literature~\cite{fraser_trust_2022,fraser_great_2024}, we select three groups of places from the 555 POI categories in our dataset: community places ($N=4,689$ across the 11 CBSAs), social businesses ($N=5,503$), and parks or public outdoor places ($N=7,860$).
As in Fraser et al.~\cite{fraser_great_2024}, we treat grocery stores, supermarkets, and post offices as baseline places ($N=6,764$), where we expect limited levels of political sorting.
To ensure comparability across cities, place-level segregation is centered on the city-specific median.
Across the 11 CBSAs in our sample, social infrastructures are, on average, visited more frequently by left-leaning individuals, in particular community places and social businesses, both with median values equal to $-0.04$ (Fig.~\ref{fig:4}b).
However, this pattern varies across metropolitan areas (SI, Fig. S4).
In some cities, such as Dallas and Miami, social infrastructures are, on average, visited by more right-leaning individuals or show lower levels of sorting, with positive median values or close to $0$.
Moreover, parks and public outdoor places in cities like Philadelphia, Los Angeles, and Washington, DC, tend to be less politically skewed than community places and social businesses.

We define visit patterns to social infrastructures as the share of total time an individual spends in each of the three categories. 
As expected, this share is relatively small compared to the time spent in other activity spaces, such as workplaces, grocery stores, and other routine venues.
% Accessibility to a given category of social places for an individual is defined as the number of points of interest in the respective category within the Census Tract where the individual lives. 
% We use Census Tract to consider POIs that are plausibly reachable, including those beyond an individual's Census Block Group (CBG).
Accessibility to a given category of social places is defined as the logarithm of the number of points of interest reachable within a 15-minute walking distance from the centroid of the individual's residential CBG, computed on the pedestrian street network.
Results are robust to alternative accessibility thresholds, including a 30-minute walking distance (see SI, Tables S7 and S8).

We estimate separate models for each city, with experienced partisan segregation as the dependent variable.
Focusing on human behavior, we consider the time spent in community, social, and public outdoor places as the main independent variables.
We also include as a control the number of places reachable within a 15-minute walking distance (i.e., our measure of local accessibility), residential partisan segregation, and the socioeconomic and demographic characteristics of the CBG of residence. 
Specifically, these include the share of graduated individuals, the shares of African American and Latino populations, income, the shares of the labor force, and the poverty rate.
See Material and Methods for a detailed description of the model specifications.

Visiting and spending time in community places and social businesses is associated with more left-leaning experienced segregation across most cities (Fig.~\ref{fig:4}a), with some exceptions such as Miami and Dallas for community places, and Miami and Philadelphia for social businesses.
Importantly, this association should be interpreted as reflecting patterns of selective use and the social composition of these places, rather than as a causal effect of the places themselves.
Specifically, time spent in community places and social businesses is predominantly associated with lower experienced partisan segregation coefficients (i.e., more left-leaning exposure) across most cities, while the association is weaker or close to zero for public outdoor POIs.
For community places, coefficients range from $-0.188$ in Boston and $-0.151$ in Philadelphia to values close to zero in Miami or Los Angeles, corresponding to differences up to $9.4\%$ and $7.6\%$ higher Democratic segregation (note that segregation takes values between -1 and +1).
Similarly, people who visit social businesses are associated with higher left-leaning experienced segregation, for example in Seattle ($8.4\%$), Boston ($6.65\%$), and San Francisco ($6.55\%$). 
Interestingly, this is not driven by local accessibility, as we control for the availability of social infrastructure within a 15-minute walking distance.
Full results of the analysis can be found in Supplementary Information, Table S7.

In general, these findings highlight the relationship between social infrastructures as central places where social capital can be built and patterns of use, suggesting that the observed partisan exposure in these settings reflects selective patterns of use and social composition rather than the effect of accessibility alone.
While the analysis does not allow for causal attribution, the findings show that visiting community places and social businesses is systematically associated with more Democratic-leaning experienced segregation across cities.

\section*{Discussion}
In the context of extensive residential partisan segregation in the United States~\cite{brown_measurement_2021,kaplan_partisan_2022,brown_sources_2025,brown_relationship_2025}, further reinforced by generational changes and ideological conformity~\cite{brown_sources_2025}, residential location remains a strong constraint on political exposure and a good proxy for the partisan environments individuals encounter in daily life, as reflected in the high correlation between residential and experienced segregation.
At the same time, residential location does not fully determine experienced partisan segregation, as evidenced by the within-area variance that remains unexplained by residence alone.
Consistent with prior works, individuals living in the same neighborhood often follow different routines~\cite{christopher_r_browning_socioeconomic_2017}, leading to different patterns of experienced partisan segregation. 
In this context, our findings show that travel range moderates, rather than mediates, the relationship between residential and experienced partisan segregation.
Similar to the mechanism behind experienced income segregation~\cite{moro_mobility_2021}, individuals who travel farther from home are exposed to more politically mixed environments, even when they live in highly segregated neighborhoods, partially escaping their residential political bubble.
However, contrary to findings on income segregation\cite{moro_mobility_2021}, after controlling for travel distance, place exploration, defined as the tendency of visiting new locations, does not reduce partisan segregation, and may even reinforce it. This difference suggests that socioeconomic and political segregation may arise through distinct behavioral mechanisms. While exploring new places can expose individuals to greater socioeconomic diversity, cross-partisan exposure may depend less on visiting unfamiliar places than on the political composition of the destinations individuals choose to visit. As a result, exploratory mobility alone may be insufficient to increase political mixing.

Moreover, consistent with activity-space analyses of partisan segregation~\cite{zhang_human_2023}, our findings confirm that place-level partisan segregation varies across activity spaces, further shaping individuals' partisan exposure, and that category-level place segregation slightly decreases with distance, similar to income segregation~\cite{moro_mobility_2021}. 
In addition, our work highlights that partisan place segregation strongly reflects the geography of political votes and is directional, with some locations attracting more left-leaning visitors and others attracting right-leaning ones.
However, not all activity spaces are equal in providing opportunities for social interaction. 
Social infrastructures are central locations that facilitate interaction and community building, fostering civic participation~\cite{klinenberg2018palaces,latham_social_2019}.
Yet, our findings suggest that community places and social businesses are, on average, visited more by left-leaning individuals across most metropolitan areas, in line with previous work~\cite{fraser_trust_2022}, but with some differences between cities.
Importantly, this pattern is not driven by local accessibility or availability, but by differences in visitation behavior.
This suggests that experienced segregation is less a function of unequal access to activity spaces and more a consequence of lifestyle in everyday routines.
From this perspective, this pattern may be associated with differences between left- and right-leaning individuals in terms of lifestyle preferences~\cite{dellaposta_why_2015,chen_effect_2018,talaifar_lifestyle_2025}, which may help explain why social infrastructures are politically skewed in most U.S. cities.
Therefore, policies aimed at fostering social interactions and social capital through social infrastructures may unintentionally reinforce partisan segregation, leading to politically homogeneous social environments.

Our study comes with limitations. 
First, our measures of partisan segregation are inferred from precinct voting patterns to capture residential exposure and, subsequently, potential exposure in activity spaces over a six-month period.
Therefore, these measures reflect contextual partisan environments rather than direct interpersonal interactions, including different forms of contact that vary in selection, depth, and duration.
However, even brief and casual encounters can impact political behavior~\cite{enos_causal_2014,enos_space_2017}.
Second, we only consider verified POIs from Foursquare, which may not capture some activity spaces or informal locations, potentially leading to measurement errors. 
Moreover, due to privacy and data agreement policies, the study does not include sensitive points of interest, such as places of worship, medical facilities, or locations associated with vulnerable populations.
These excluded locations (e.g., places of worship) may function as community spaces for specific groups of individuals and have a relatively homogeneous political composition.
Lastly, to enhance data representativeness, we apply a re-weighting based on the sample distribution across Census Block Groups, which does not account for individual-level demographic and socioeconomic characteristics.

Despite these limitations, our research advances the understanding of experienced political segregation from a behavioral perspective using granular anonymized individual mobility data. While residential context remains a strong baseline for political exposure, it does not fully determine the political environments individuals experience in their daily lives. Instead, experienced partisan segregation also emerges from the ways people move through cities and the destinations they choose to visit. Individuals living in similar residential environments can therefore experience markedly different political environments depending on their everyday mobility routines. Travel range enables individuals to partially escape the constraints of residential political segregation, whereas simply exploring a wider variety of places does not. Together, these findings show that experienced partisan segregation is not only a property of residential geography, but also of everyday behavior.

\section*{Materials and methods}

\subsection*{Data sources}
Our research integrates high-resolution anonymized individual mobility data provided by Cuebiq, collected between October 2016 and April 2017, with presidential vote results at the precinct level from the 2016 election, as well as socioeconomic data from the Census at the Census Block Group (CBG) level.

Mobility data consists of individuals in Cuebiq's panel who opted-in and provided informed consent for data collection, in compliance with the General Data Protection Regulation (GDPR) and the California Consumer Privacy Act (CCPA).

Specifically, mobility data includes $1,219,439$ individuals in 11 major U.S. Core-Based Statistical Areas (CBSAs) (i.e., Boston, Chicago, Dallas, Detroit, Los Angeles, Miami, New York, Philadelphia, San Francisco, Seattle, Washington DC; see SI, Table S1), after applying the filtering procedure described in SI, Section S1.
Stop locations are defined as locations in which a user stops for at least 5 minutes.
Each stop location is then assigned to the nearest POI within a defined distance threshold, using the default Foursquare taxonomy, resulting in 555 non-sensitive activity space categories.
We enhance data representativeness by applying post-stratification re-weighting based on population distributions across CBGs, both in the computation of segregation and in subsequent modeling.
As in Moro et al. (2021)\cite{moro_mobility_2021}, weights are computed by scaling observations according to the population of the Census Block Group (CBG) of residence. 
Specifically, each individual is assigned a weight proportional to the ratio between the total population of the CBG and the number of sampled individuals in that CBG.
Further details of data pre-processing can be found in the Supplementary Information of Moro et al. (2021)~\cite{moro_mobility_2021}.
The use of the data in this research was reviewed by the Northeastern University Institutional Review Board (IRB \#24-03-43).

Precinct-level voting results from the 2016 U.S. presidential election were obtained from the Voting and Election Science Team through the Harvard Dataverse~\cite{DVN/NH5S2I_2018}. Census data are retrieved from the 2012-2016 5-year American Community Survey at the Census Block Group (CBG) level~\cite{uscensusbureau2017acs}. 
Specifically, we consider the share of individuals with a bachelor's or higher degree, the shares of Hispanic or Latino and African American residents, the poverty rate, labor force participation, and median household income.
Drawing on the literature on social segregation, on the one hand, and political behavior, on the other, these variables capture the core socioeconomic and demographic cleavages associated with residential sorting and political alignment in the United States.
Educational attainment is a key predictor of political preferences and geographic clustering, with higher-educated individuals concentrated in urban areas and more likely to support Democratic candidates~\cite{brown_sequential_2024}.
Race and ethnicity are central to both residential segregation~\cite{intrator_segregation_2016} and voting behavior~\cite{kuriwaki_geography_2024}, reflecting long-standing patterns of spatial inequality.
Finally, income inequality and economic indicators are strongly associated with residential segregation~\cite{reardon_income_2011}.

\subsection*{Classification of points of interest and social infrastructures}
As mentioned, activity spaces are categorized using the default Foursquare taxonomy, resulting in 555 categories of points of interest (POIs) in our dataset.
Drawing on the literature on social infrastructure and social capital~\cite{fraser_trust_2022,fraser_great_2024}, we manually categorize each POI into community places, social businesses, and parks or public outdoor spaces to examine the relationship between experienced partisan segregation and visitation patterns to social infrastructures.
Community places include locations such as libraries, art galleries, museums, arts, academic, and administrative buildings. 
These venues are typically civic, cultural, or educational places that facilitate repeated non-commercial social interaction.
Social businesses consist of commercial venues such as bars, cafés, bookstores, or coworking spaces. 
These locations are characterized by a monetary cost of entry or consumption, with social interaction occurring alongside commercial activity.
Finally, parks and public outdoor spaces include publicly accessible environments such as parks, sidewalks, sports venues, swimming pools, and other recreational facilities that support unstructured or leisure-oriented interaction.
As a baseline category, we consider grocery stores, post offices, and supermarkets as locations for daily necessities where we do not expect consistent and unidirectional partisan sorting.
The full list of social infrastructures can be found in the Supplementary Information, Table S6.

\subsection*{Individual and place segregation}
First, we compute the political composition of visitors for each point of interest (POI) based on the time individuals spent at each location and their inferred political affiliation.
Political affiliation is inferred from the results of the 2016 presidential election in the precinct where each individual resides, defined as the difference between the share of votes for Republicans and Democrats.
This measure ranges from $-1$ (if the entire precinct voted for Democrats) to $1$ (if the entire precinct voted for Republicans).
In accordance with Cuebiq’s Sensitive Points of Interest (SPOI) policy, we exclude sensitive points of interest, such as healthcare-related sites, child facilities, religious buildings, and places associated with minority and vulnerable populations.
Place segregation, $S_{\alpha}$, is defined as the time-weighted average of the residential partisan segregation of all visitors as follows.

\begin{equation}
S_{\alpha} = \frac{\sum_{i=1}^{N} t_{i, \alpha} S_{i}^{res}}{\sum_{i=1}^{N} t_{i, \alpha}}
\end{equation}

Second, we calculate each user's experienced segregation based on the political composition of the POIs they visit, not considering the self-contribution to the place segregation using a leave-one-out measure.
Similar to political affiliation, both place- and individual-level experienced segregation range from –1 (Democratic segregation) to 1 (Republican segregation).
Individual political segregation, $S_{i}^{exp}$, is defined as follows:

% \begin{equation}
% S_{i}^{exp} = \frac{\sum_{\alpha=1}^{N} t_{i, \alpha} S_{\alpha}}{\sum_{\alpha=1}^{N} t_{i, \alpha}}
% \end{equation}

\begin{equation}
S_{i}^{exp} = \frac{\sum_{\alpha=1}^{N} t_{i, \alpha} S_{\alpha}^{(-i)}}{\sum_{\alpha=1}^{N} t_{i, \alpha}}
\end{equation}

\subsection*{Multilevel regression models for variance decomposition}
We measure within-area variance in place-level and individual-level political segregation using multilevel regression models with two nested geographical levels: U.S. Census Tracts and Counties. 
To this end, for each city, we model segregation with random intercepts for counties and tracts and compute the Variance Partition Coefficient (VPC) as the proportion of total variance that is attributable to each hierarchical level in the multilevel model.

\begin{equation}
S_{ijk} = \beta_0 + u_k + v_{jk} + \varepsilon_{ijk}
\end{equation}

where $u_k \sim N(0, \sigma^2_{\text{county}})$ is the random intercept for counties, $v_{jk} \sim N(0, \sigma^2_{\text{tract}})$ is the random intercept for tracts, and $\varepsilon_{ijk} \sim N(0, \sigma^2_{\text{residual}})$ is the residual variance.

\subsection*{Mobility behavior measures}
We examine mobility behavior using two distinct metrics.
First, we consider the average distance traveled, $\bar{d}_i$, as the logarithm of the time-weighted average distance (in kilometers) between an individual's home location and the points of interest they visit.
This metric is particularly relevant given the strong spatial autocorrelation of political votes in U.S. metropolitan areas, where individuals are clustered in politically homogeneous neighborhoods, especially in central urban cores and more peripheral areas.
Second, we compute the place exploration rate, $\sigma_p$, for each individual, defined as the probability of visiting a new location. 
This metric characterizes individual mobility behavior along the explorer-returner dichotomy~\cite{cuttone_understanding_2018,pappalardo_returners_2015}, reflecting the tendency to explore new places or repeatedly visit familiar ones.

\begin{equation}
\sigma_{p,i} = \frac{unique(N_{\alpha,i})}{N_{\alpha,i}}
\end{equation}

where $N_{\alpha,i}$ is the total number of visits to POIs, $\alpha$, of an individual $i$ and $unique(N_{\alpha,i})$ is the distinct number of POIs visited.

\subsection*{Multilevel regression models for individual political segregation}
We investigate how mobility behavior moderates residential segregation in the relationship with individual political segregation using multilevel regression models. 
We use experienced partisan segregation, $S_{i}^{exp}$, as dependent variable.
To test the moderation effect of mobility behavior, we introduce two interaction terms between residential partisan segregation, $S_{i}^{res}$ and mobility metrics, specifically average distance traveled, $\bar{d_{i}}$ and place exploration, $\sigma_{p,i}$.
We control for socioeconomic and demographic characteristics at the CBG level, accounting for the share of graduated individuals, the shares of African Americans and Latinos, income, and the shares of labor force and poverty rate.
The model specification is the following.

\begin{equation}
S_{i}^{exp} =
\beta_0
+ \beta_1 S_{i}^{res}
+ \boldsymbol{\beta}_2^{\top} \mathbf{X}_i
+ \beta_3 \sigma_{p,i}
+ \beta_4 \bar{d}_i
+ \beta_5 \left( S_{i}^{res} \times \sigma_{p,i} \right)
+ \beta_6 \left( S_{i}^{res} \times \bar{d}_i \right)
+ u_{ct} + v_c + \epsilon_i
\end{equation}

where $ \mathbf{X}_i$ represents the vector of control variables previously mentioned.
We use cities and counties as two nested levels for the model that consider all the metropolitan areas together, and counties and tracts for the single-city models.

\subsection*{Models for social infrastructures and experienced segregation}
To examine the relationship between experienced partisan segregation and visitation patterns to social infrastructures, we estimate linear regression models separately for each metropolitan area.

We use individual experienced partisan segregation, $S_{i}^{exp}$, as the dependent variable. 
Our main independent variables capture behavioral exposure, defined as the share of total time an individual spends in community places, social businesses, and public outdoor spaces.
% To account for structural opportunity, we include category-specific accessibility measures, defined as the number of points of interest in each category within the Census Tract of residence (log-transformed). 
To account for structural opportunity, we include category-specific accessibility measures, defined as the logarithm of the number of points of interest in each category reachable within a 15-minute walking distance from the centroid of the individual's residential CBG. 
We also control for residential partisan segregation, $S_{i}^{res}$, and for socioeconomic and demographic characteristics at the Census Block Group level.

The model specification is:
\begin{equation}
S_{i}^{exp} =
\beta_0
+ \beta_1 S_{i}^{res}
+ \beta_2 \text{c}_i
+ \beta_3 \text{sb}_i
+ \beta_4 \text{o}_i
+ \beta_5 \log(N_{\text{c},i})
+ \beta_6 \log(N_{\text{sb},i})
+ \beta_7 \log(N_{\text{o},i})
+ \boldsymbol{\beta}_8^{\top} \mathbf{X}_i
+ \epsilon_i
\end{equation}

where $\text{c}_i$, $\text{sb}_i$, and $\text{o}_i$ denote the share of time spent in each category, and $N_{\text{c},i}$, $N_{\text{sb},i}$, and $N_{\text{o},i}$ represent the number of corresponding points of interest.

\section*{Data and code availability}
Precinct-level data on the 2016 U.S. Presidential election are publicly available through the Harvard Dataverse~\cite{DVN/NH5S2I_2018}.
Mobility data are available upon application to Cuebiq's Social Impact program (https://cuebiq.com/social-impact/) and were used under license for the current study.
Researchers may request access directly from Cuebiq through its Social Impact program, subject to Cuebiq's eligibility requirements and data-use conditions (\url{https://cuebiq.com/social-impact}, contact \url{https://cuebiq.com/contact/}). 
Data and code to reproduce the main results of the study are publicly available at \url{https://github.com/tonmarco/urban-partisan-segregation}.

\section*{Acknowledgements}
M.To. acknowledges the support of the NRRP MUR program funded by the NextGenerationEU. 
The work of B.L. was partially supported by the following projects: Horizon Europe Programme, grant \#101120237-ELIAS and grant \#101120763-TANGO. Funded by the European Union. Views and opinions expressed are however those of the author(s) only and do not necessarily reflect those of the European Union or the European Health and Digital Executive Agency (HaDEA). Neither the European Union nor the granting authority can be held responsible for them.
E.M. acknowledges the support of NSF grant 2420945.

\section*{Author contributions}
M.To. analyzed the data, performed the research, and drafted the manuscript.
M.To., M.Ti., B.L., E.M. designed the study, interpreted the results, provided critical feedback, and revised the manuscript.
All authors approved the final version of the manuscript.

\section*{Competing interests}
The authors declare no competing interests.

\bibliography{sample}

\end{document}

\title{Supplementary Information of: Experienced partisan segregation across patterns of mobility behavior in US cities}

% Authorship
\author{Marco Tonin}
\affiliation{Department of Sociology and Social Research, University of Trento, Trento, Italy}
\affiliation{Fondazione Bruno Kessler (FBK), Trento, Italy}

\author{Michele Tizzoni}
\affiliation{Department of Sociology and Social Research, University of Trento, Trento, Italy}

\author{Bruno Lepri}
\affiliation{Fondazione Bruno Kessler (FBK), Trento, Italy}

\author{Esteban Moro}
\affiliation{Network Science Institute, Department of Physics, Northeastern University, Boston, 02115, MA, USA}

%\thanks{These authors contributed equally to this work.}

\maketitle
\tableofcontents

\newpage
%%%%%%%%%%%%%%%%%%%%%%%%% ---- %%%%%%%%%%%%%%%%%%%%%%%
\section{Data sources}
\label{SI:sec:datasets}

Mobility data includes individuals in the Cuebiq panel who opted-in and provided informed consent for data collection, in compliance with the General Data Protection Regulation (GDPR) and the California Consumer Privacy Act (CCPA).
To enhance data quality, we apply several filters, which result in the final sample outlined in Table \ref{tab:msa}.
First, we exclude sensitive points of interest, such as healthcare-related sites, child facilities, religious
buildings, and places related to minorities, reducing the number of visits of the individuals in the sample.
Second, we define a visit to a point of interest  as a stop location that is located more than 50 meters from the individuals' home and less than 75 meters from the centroid of the POI.
Third, we consider only points of interest with at least 10 unique visitors and individuals with at least 5 unique POIs visited, to ensure minimal behavioral coverage while preserving sample size.
These filtering steps result in a final sample of 1,219,439 individuals.

\begin{table}[h]
\centering
\begin{tabular}{l c cc}
\hline
 &  & \multicolumn{2}{c}{\textbf{Sample filtered}} \\
\textbf{City (State)} & \textbf{Population} & \textbf{\# Devices} & \textbf{\# Stays} \\
\hline

Boston-Cambridge-Newton (MA-NH) & 4.64 M & 40,124  &  1,645,674 \\
Chicago-Naperville-Elgin (IL-IN-WI) & 9.52 M & 190,213  & 8,352,230 \\
Dallas-Fort Worth-Arlington (TX) & 6.70 M & 169,326  &  8,004,874 \\
Detroit-Warren-Dearborn (MI) & 4.29 M & 72,857  & 3,086,558 \\
Los Angeles-Long Beach-Anaheim (CA) & 13.05 M & 176,903 & 7,988,619 \\
Miami-Fort Lauderdale-West Palm Beach (FL) & 5.76 M & 127,343 & 5,619,352 \\
New York-Jersey City (NY-NJ-PA) & 19.83 M & 158,261 & 7,107,880 \\
Philadelphia-Camden-Wilmington (PA-NJ-DE-MD) & 6.02 M & 76,921 & 3,055,496 \\
San Francisco-Oakland-Hayward (CA) & 4.46 M & 56,023 & 2,094,449 \\
Seattle-Tacoma-Bellevue (WA) & 3.55 M & 50,184 & 2,016,236 \\
Washington-Arlington-Alexandria (DC-VA-MD-WV) & 5.86 M & 101,284 & 4,211,369 \\

\hline
\textbf{Total} & 83.5 M & 1,219,439 & 53,182,737 \\
\hline

\end{tabular}
\caption{Summary statistics of devices and stays across metropolitan areas}
\label{tab:msa}
\end{table}

\clearpage

\section{Variance decomposition for experienced segregation}
\label{SI:sec:variance}

As described in the manuscript, we estimate within-areas variance of experienced partisan segregation using multilevel regression models with random intercepts nested at two geographical levels (county and census tract).

\begin{figure*}[ht!]
\centering
\includegraphics[width=0.7\textwidth]{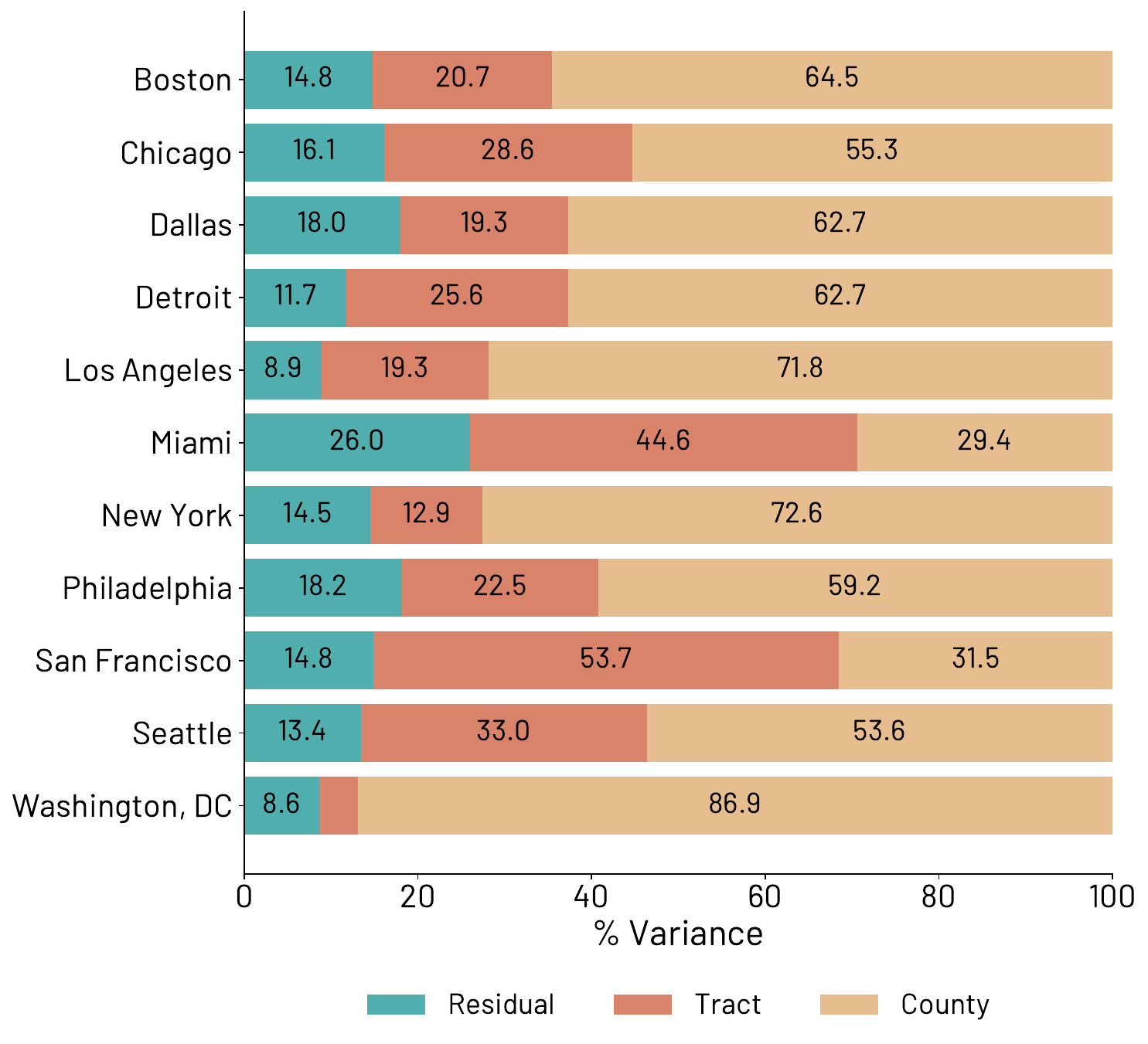} 
\caption{\textbf{Variance Partition Coefficients (VPCs) of experienced partisan segregation.} Variance Partition Coefficients (VPCs) from the multilevel regression models show moderate heterogeneity of experienced partisan segregation that is not explained by differences between Counties and Tracts}
\label{fig:SI_vpc}
\end{figure*}

As in the analysis of place partisan segregation, differences between counties and tracts account for most of the variance, with within-areas variance ranging from $8.6\%$ in Washington, DC to $26\%$ in Miami, as outlined in Fig. \ref{fig:SI_vpc}.

\clearpage

\section{Mobility behavior as mediator or moderator between residential and experienced partisan segregation}
\label{SI:sec:mediation}
\begin{figure*}[!ht!]
\centering
\includegraphics[width=1\textwidth]{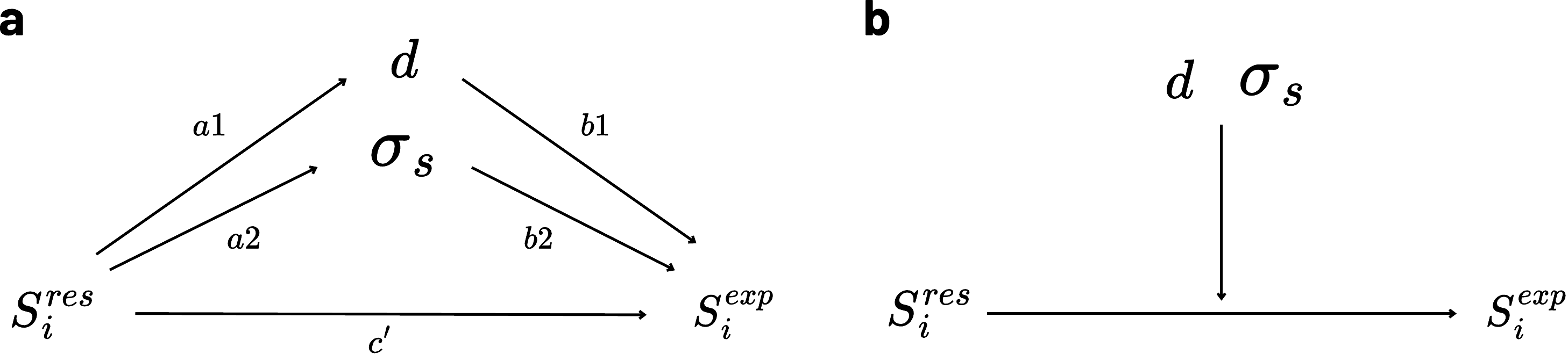} 
\caption{\textbf{Mediation and moderation diagrams} \textbf{(a)} Hypothesis on the mediation effect of mobility behavior in the relationship between residential and experienced segregation. Here, $a1, a2, b1$ and $b2$ represent the indirect paths of the relationship between residential and experienced partisan segregation, while $c'$ represents the direct effect; \textbf{(b)} Hypothesis on the moderation effect of mobility behavior in the relationship between residential and experienced segregation, with mobility behavior that impact the effect of residential segregation on the experienced partisan segregation.}
\label{fig:SI_diag}
\end{figure*}

We investigate the role of mobility behavior in the relationship between residential and experienced partisan segregation by testing whether it acts as a mediator or a moderator.
In the former case, residential segregation influences how individuals move within the urban context, which in turn shapes their experienced segregation, whereas in the latter mobility behavior moderates the strength of this relationship.
This is described by Figure \ref{fig:SI_diag}, which presents the conceptual relationships as diagrams.

As reported in the manuscript and in Table \ref{tab:mediation}, mobility behavior acts as a moderator rather than a mediator of the relationship between residential and experienced partisan segregation, with effects that vary depending on individuals’ place of residence.

\begin{figure*}[ht!]
\centering
\includegraphics[width=1\textwidth]{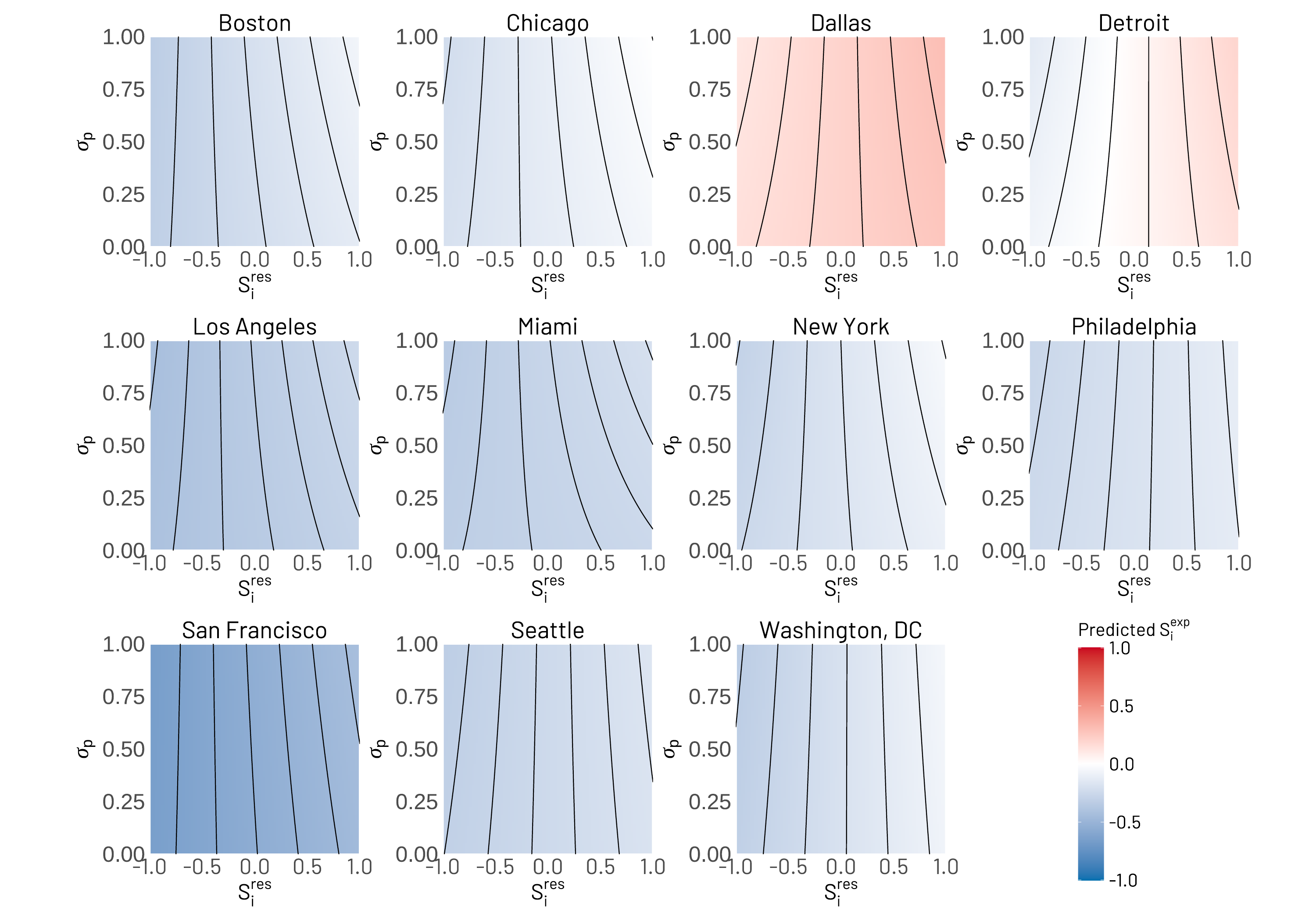} 
\caption{\textbf{Moderation effects of exploration rate across the 11 metropolitan areas.}}
\label{fig:SI_contour}
\end{figure*}

Fig. \ref{fig:SI_contour} shows the moderating effect of the exploration rate metric, defined as the tendency of visiting new places.
Finally, Tables \ref{tab:performances}, \ref{tab:regressions}, and \ref{tab:regressions_scaled} report the performances of the models with and without mobility metrics using $R^2$ and $BIC$, which penalizes when adding new independent variables.

\vspace{1cm}

\begin{table*}[ht!]
\centering
\caption{\textbf{Mediation analysis results by metropolitan area.}
$a_1$ and $a_2$ represent the effects of relative residential segregation on mobility measures, 
$b_1$ and $b_2$ represent the effects of mobility on experienced segregation, 
$\text{ind}_1$ and $\text{ind}_2$ are the indirect effects through each mobility pathway, 
$\text{ind}_{\text{total}}$ is the total indirect effect, and $c'$ is the direct effect.
Here, direct effects are much larger than the indirect effects.}
\label{tab:mediation}

\begin{tabular}{lrrrrrrrr}
CBSA & $a_1$ & $a_2$ & $b_1$ & $b_2$ & $\text{ind}_1$ & $\text{ind}_2$ & $\text{ind}_{\text{total}}$ & $c'$ \\
\midrule
Boston
& 0.643 & -0.00966 & -0.00884 & 0.00958 & -5.68e-03 & -9.25e-05 & -5.77e-03 & 0.167 \\
Chicago
& 0.214 & 0.0111 & -0.00368 & -0.00232 & -7.90e-04 & -2.58e-05 & -8.16e-04 & 0.0903 \\
Dallas
& 0.0736 & -0.00139 & -0.00363 & -0.00603 & -2.68e-04 & 8.36e-06 & -2.59e-04 & 0.0642 \\
Detroit
& 0.0540 & -0.0204 & -0.000263 & -0.0213 & -1.42e-05 & 4.33e-04 & 4.19e-04 & 0.158 \\
Los Angeles
& 0.306 & -0.0162 & 0.00329 & -0.000634 & 1.01e-03 & 1.03e-05 & 1.02e-03 & 0.0807 \\
Miami
& 0.0677 & 0.0148 & 0.00304 & 0.00322 & 2.06e-04 & 4.78e-05 & 2.54e-04 & 0.0483 \\
New York
& 0.168 & -0.00592 & -0.0100 & -0.0111 & -1.68e-03 & 6.58e-05 & -1.62e-03 & 0.125 \\
Philadelphia
& 0.214 & 0.00565 & 0.0106 & -0.0175 & 2.26e-03 & -9.87e-05 & 2.17e-03 & 0.0809 \\
San Francisco
& 0.177 & -0.0185 & 0.00103 & -0.00224 & 1.82e-04 & 4.14e-05 & 2.24e-04 & 0.0715 \\
Seattle
& 0.440 & -0.0163 & 0.000370 & -0.0139 & 1.63e-04 & 2.25e-04 & 3.88e-04 & 0.0785 \\
Washington, DC
& 0.406 & -0.00909 & 0.00902 & -0.0184 & 3.66e-03 & 1.67e-04 & 3.83e-03 & 0.0933 \\
\bottomrule
\end{tabular}
\end{table*}

\clearpage

\begin{table}[h]
\centering
\begin{tabular}{llccc}
\hline
\textbf{CBSA} & \textbf{Model specification} & \textbf{Marginal $R^2$} & \textbf{Conditional $R^2$} & \textbf{BIC} \\
\hline

Boston & $S_{i}^{res}$ & 0.113 & 0.779 & -80902.69 \\
       & $S_{i}^{res}$ + controls & 0.132 & 0.777 & -80870.69 \\
       & $S_{i}^{res}$ + controls + mobility & 0.132 & 0.851 & -89348.74 \\
\hline

Chicago & $S_{i}^{res}$ & 0.044 & 0.780 & -311187.6 \\
        & $S_{i}^{res}$ + controls & 0.070 & 0.771 & -311516.0 \\
        & $S_{i}^{res}$ + controls + mobility & 0.103 & 0.874 & -357577.1 \\
\hline

Dallas & $S_{i}^{res}$ & 0.022 & 0.779 & -249434.6 \\
       & $S_{i}^{res}$ + controls & 0.032 & 0.779 & -249773.8 \\
       & $S_{i}^{res}$ + controls + mobility & 0.074 & 0.866 & -289594.4 \\
\hline

Detroit & $S_{i}^{res}$ & 0.134 & 0.798 & -92652.78 \\
        & $S_{i}^{res}$ + controls & 0.154 & 0.798 & -92906.83 \\
        & $S_{i}^{res}$ + controls + mobility & 0.132 & 0.880 & -110413.8 \\
\hline

Los Angeles & $S_{i}^{res}$ & 0.018 & 0.885 & -377640.3 \\
            & $S_{i}^{res}$ + controls & 0.024 & 0.878 & -377766.7 \\
            & $S_{i}^{res}$ + controls + mobility & 0.051 & 0.907 & -415708.8 \\
\hline

Miami & $S_{i}^{res}$ & 0.028 & 0.690 & -296098.0 \\
      & $S_{i}^{res}$ + controls & 0.052 & 0.678 & -296501.1 \\
      & $S_{i}^{res}$ + controls + mobility & 0.107 & 0.716 & -311980.7 \\
\hline

New York & $S_{i}^{res}$ & 0.093 & 0.792 & -185717.5  \\
         & $S_{i}^{res}$ + controls & 0.105 & 0.791 & -186089.9 \\
         & $S_{i}^{res}$ + controls + mobility & 0.134 & 0.865 & -222410.3 \\
\hline

Philadelphia & $S_{i}^{res}$ & 0.050 & 0.756 & -112813.2 \\
             & $S_{i}^{res}$ + controls & 0.063 & 0.753 & -112818.5 \\
             & $S_{i}^{res}$ + controls + mobility & 0.128 & 0.832 & -128588.7 \\
\hline

San Francisco & $S_{i}^{res}$ & 0.024 & 0.819 & -139936.9 \\
              & $S_{i}^{res}$ + controls & 0.027 & 0.819 & -139831.8 \\
              & $S_{i}^{res}$ + controls + mobility & 0.139 & 0.855 & -153443.8 \\
\hline

Seattle & $S_{i}^{res}$ & 0.017 & 0.834 & -80992.77 \\
        & $S_{i}^{res}$ + controls & 0.033 & 0.822 & -80990.77 \\
        & $S_{i}^{res}$ + controls + mobility & 0.113 & 0.882 & -96017.04 \\
\hline

Washington, DC & $S_{i}^{res}$ & 0.024 & 0.888 & -143208.0 \\
              & $S_{i}^{res}$ + controls & 0.027 & 0.887 & -143192.9 \\
              & $S_{i}^{res}$ + controls + mobility & 0.076 & 0.923 & -169436.3 \\
\hline

\end{tabular}
\caption{Model fit statistics across metropolitan areas. Including mobility metrics, model performances improve according the conditional $R^2$ and BIC metric, which penalizes the larger number of independent variables.}
\label{tab:performances}
\end{table}

\clearpage
\begin{table*}[!htbp]
\centering
\caption{Mixed-effects model results by city}
\resizebox{\textwidth}{!}{%
\begin{tabular}{lccccccccccc}
\toprule
& \multicolumn{11}{c}{\textbf{Metropolitan Areas}} \\
\cmidrule(lr){2-12}
& BOS & CHI & DAL & DET & LA & MIA & NYC & PHI & SF & SEA & DC \\
\midrule
$S_{res}$
& 0.488*** & 0.517*** & 0.513*** & 0.650*** & 0.419*** & 0.284*** & 0.452*** & 0.467*** & 0.462*** & 0.575*** & 0.591*** \\
& (0.007) & (0.003) & (0.004) & (0.007) & (0.003) & (0.003) & (0.003) & (0.005) & (0.005) & (0.006) & (0.005) \\
$\sigma_p$
& 0.024*** & 0.014*** & -0.002* & -0.009*** & 0.014*** & 0.016*** & 0.009*** & -0.006** & 0.012*** & -0.001 & -0.002 \\
& (0.002) & (0.001) & (0.001) & (0.002) & (0.001) & (0.001) & (0.001) & (0.002) & (0.003) & (0.002) & (0.002) \\
$\bar{d}_i$
& -0.072*** & -0.071*** & -0.011*** & -0.031*** & -0.068*** & -0.033*** & -0.068*** & -0.049*** & -0.108*** & -0.083*** & -0.087*** \\
& (0.001) & (0.000) & (0.000) & (0.001) & (0.000) & (0.000) & (0.000) & (0.001) & (0.001) & (0.001) & (0.001) \\
\% Graduates
& -0.027*** & -0.008* & 0.012*** & 0.001 & 0.001 & 0.026*** & -0.002 & -0.015** & -0.006 & -0.024*** & -0.003 \\
& (0.005) & (0.003) & (0.004) & (0.007) & (0.003) & (0.003) & (0.004) & (0.005) & (0.003) & (0.005) & (0.004) \\
\% African Americans
& -0.026*** & -0.075*** & -0.040*** & -0.092*** & -0.027*** & -0.037*** & -0.065*** & -0.040*** & -0.011* & 0.018* & -0.040*** \\
& (0.007) & (0.003) & (0.003) & (0.005) & (0.004) & (0.003) & (0.003) & (0.004) & (0.005) & (0.007) & (0.004) \\
\% Hispanic
& 0.001 & -0.021*** & -0.037*** & -0.018 & -0.017*** & -0.001 & -0.042*** & -0.006 & -0.005 & 0.003 & -0.022*** \\
& (0.006) & (0.003) & (0.003) & (0.011) & (0.002) & (0.003) & (0.003) & (0.006) & (0.003) & (0.005) & (0.004) \\
Log Income
& 0.005* & -0.002 & 0.000 & 0.002 & 0.004*** & 0.002* & 0.001 & 0.001 & 0.003* & -0.003 & -0.005** \\
& (0.002) & (0.001) & (0.001) & (0.002) & (0.001) & (0.001) & (0.001) & (0.002) & (0.001) & (0.002) & (0.002) \\
\% Poverty Ratio
& -0.012 & -0.010** & -0.006 & -0.019** & -0.010*** & -0.007* & 0.006 & -0.011* & 0.011* & 0.009 & 0.010 \\
& (0.007) & (0.003) & (0.004) & (0.006) & (0.003) & (0.003) & (0.005) & (0.006) & (0.004) & (0.006) & (0.006) \\
\% Labor Force
& 0.001 & 0.005 & -0.002 & 0.014* & -0.004 & -0.008** & 0.002 & -0.002 & -0.002 & -0.004 & 0.001 \\
& (0.006) & (0.003) & (0.004) & (0.006) & (0.003) & (0.003) & (0.004) & (0.005) & (0.003) & (0.005) & (0.005) \\
$S_{res} \times \sigma_p$
& 0.043*** & 0.043*** & 0.038*** & 0.064*** & 0.037*** & 0.041*** & 0.053*** & 0.019*** & 0.023*** & 0.018*** & 0.024*** \\
& (0.005) & (0.002) & (0.002) & (0.004) & (0.002) & (0.003) & (0.003) & (0.004) & (0.005) & (0.005) & (0.004) \\
$S_{res} \times \bar{d}_i$
& -0.179*** & -0.191*** & -0.190*** & -0.236*** & -0.165*** & -0.117*** & -0.167*** & -0.183*** & -0.176*** & -0.226*** & -0.204*** \\
& (0.002) & (0.001) & (0.001) & (0.002) & (0.001) & (0.001) & (0.001) & (0.001) & (0.001) & (0.002) & (0.001) \\
\midrule
\multicolumn{12}{l}{\textbf{Model Fit}} \\
N
& 40,124 & 190,213 & 169,326 & 72,857 & 176,903 & 127,343 & 158,261 & 76,921 & 56,023 & 50,184 & 101,284 \\
R$^2$ (marginal)
& 0.132 & 0.103 & 0.074 & 0.132 & 0.051 & 0.107 & 0.134 & 0.129 & 0.139 & 0.113 & 0.076 \\
R$^2$ (conditional)
& 0.851 & 0.874 & 0.866 & 0.880 & 0.907 & 0.716 & 0.865 & 0.832 & 0.855 & 0.882 & 0.923 \\
BIC
& -89,348.7 & -357,577.1 & -289,594.4 & -110,413.8 & -415,708.8 & -311,980.7 & -222,410.3 & -128,588.7 & -153,443.8 & -96,017 & -169,436.3 \\
\bottomrule
\multicolumn{12}{l}{\footnotesize \textit{Note:} BOS = Boston, CHI = Chicago, DAL = Dallas, DET = Detroit, LA = Los Angeles,} \\
\multicolumn{12}{l}{\footnotesize MIA = Miami, NYC = New York, PHI = Philadelphia, SF = San Francisco, SEA = Seattle, DC = Washington, DC.} \\
\multicolumn{12}{l}{\footnotesize Standard errors in parentheses. * p$<$0.05, ** p$<$0.01, *** p$<$0.001.} \\
\end{tabular}
}
\label{tab:regressions}
\end{table*}

\clearpage

\clearpage
\begin{table*}[!htbp]
\centering
\caption{Mixed-effects model results by city (standardized coefficients)}
\resizebox{\textwidth}{!}{%
\begin{tabular}{lccccccccccc}
\toprule
& \multicolumn{11}{c}{\textbf{Metropolitan Areas}} \\
\cmidrule(lr){2-12}
& BOS & CHI & DAL & DET & LA & MIA & NYC & PHI & SF & SEA & DC \\
\midrule
$S_{res}$
& 0.035*** & 0.035*** & 0.035*** & 0.055*** & 0.026*** & 0.018*** & 0.049*** & 0.024*** & 0.023*** & 0.021*** & 0.045*** \\
& (0.002) & (0.001) & (0.001) & (0.002) & (0.001) & (0.001) & (0.001) & (0.001) & (0.001) & (0.001) & (0.001) \\
$\sigma_p$
& 0.003*** & 0.001** & -0.000 & -0.003*** & -0.000 & 0.001*** & -0.001** & -0.002*** & -0.000 & -0.001*** & -0.003*** \\
& (0.000) & (0.000) & (0.000) & (0.000) & (0.000) & (0.000) & (0.000) & (0.000) & (0.000) & (0.000) & (0.000) \\
$\bar{d}_i$
& -0.017*** & -0.016*** & -0.012*** & -0.012*** & 0.001** & -0.001*** & -0.026*** & -0.002*** & -0.002*** & -0.007*** & 0.003*** \\
& (0.000) & (0.000) & (0.000) & (0.000) & (0.000) & (0.000) & (0.000) & (0.000) & (0.000) & (0.000) & (0.000) \\
\% Graduates
& -0.006*** & -0.002* & 0.003*** & 0.000 & 0.000 & 0.005*** & -0.000 & -0.003** & -0.001 & -0.005*** & -0.001 \\
& (0.001) & (0.001) & (0.001) & (0.001) & (0.001) & (0.001) & (0.001) & (0.001) & (0.001) & (0.001) & (0.001) \\
\% African Americans
& -0.003*** & -0.018*** & -0.007*** & -0.023*** & -0.003*** & -0.008*** & -0.014*** & -0.009*** & -0.001* & 0.001* & -0.010*** \\
& (0.001) & (0.001) & (0.001) & (0.001) & (0.000) & (0.001) & (0.001) & (0.001) & (0.000) & (0.001) & (0.001) \\
\% Hispanic
& 0.000 & -0.004*** & -0.008*** & -0.001 & -0.005*** & -0.000 & -0.008*** & -0.001 & -0.001 & 0.000 & -0.003*** \\
& (0.001) & (0.001) & (0.001) & (0.001) & (0.001) & (0.001) & (0.001) & (0.001) & (0.000) & (0.001) & (0.001) \\
Log Income
& 0.002* & -0.001 & 0.000 & 0.001 & 0.002*** & 0.001* & 0.000 & 0.001 & 0.001* & -0.001 & -0.002** \\
& (0.001) & (0.001) & (0.001) & (0.001) & (0.000) & (0.001) & (0.001) & (0.001) & (0.001) & (0.001) & (0.001) \\
\% Poverty Ratio
& -0.001 & -0.001** & -0.001 & -0.003** & -0.001*** & -0.001* & 0.001 & -0.001* & 0.001* & 0.001 & 0.001 \\
& (0.001) & (0.000) & (0.000) & (0.001) & (0.000) & (0.000) & (0.001) & (0.001) & (0.000) & (0.001) & (0.001) \\
\% Labor Force
& 0.000 & 0.000 & -0.000 & 0.001* & -0.000 & -0.001** & 0.000 & -0.000 & -0.000 & -0.000 & 0.000 \\
& (0.001) & (0.000) & (0.000) & (0.001) & (0.000) & (0.000) & (0.000) & (0.001) & (0.000) & (0.001) & (0.000) \\
$S_{res} \times \sigma_p$
& 0.003*** & 0.003*** & 0.004*** & 0.006*** & 0.003*** & 0.003*** & 0.005*** & 0.001*** & 0.001*** & 0.001*** & 0.002*** \\
& (0.000) & (0.000) & (0.000) & (0.000) & (0.000) & (0.000) & (0.000) & (0.000) & (0.000) & (0.000) & (0.000) \\
$S_{res} \times \bar{d}_i$
& -0.042*** & -0.051*** & -0.056*** & -0.059*** & -0.038*** & -0.025*** & -0.063*** & -0.046*** & -0.031*** & -0.048*** & -0.055*** \\
& (0.000) & (0.000) & (0.000) & (0.000) & (0.000) & (0.000) & (0.000) & (0.000) & (0.000) & (0.000) & (0.000) \\
\midrule
\multicolumn{12}{l}{\textbf{Model Fit}} \\
N
& 40,124 & 190,213 & 169,326 & 72,857 & 176,903 & 127,343 & 158,261 & 76,921 & 56,023 & 50,184 & 101,284 \\
R$^2$ (marginal)
& 0.132 & 0.103 & 0.074 & 0.132 & 0.051 & 0.107 & 0.134 & 0.127 & 0.139 & 0.113 & 0.076 \\
R$^2$ (conditional)
& 0.851 & 0.874 & 0.866 & 0.880 & 0.907 & 0.716 & 0.865 & 0.832 & 0.855 & 0.882 & 0.923 \\
BIC
& -89,312.8 & -357,544.3 & -289,562.2 & -110,379.0 & -415,850.9 & -311,947.6 & -222,379.3 & -128,554.5 & -153,406.2 & -95,979.2 & -169,402.3 \\
\bottomrule
\multicolumn{12}{l}{\footnotesize \textit{Note:} BOS = Boston, CHI = Chicago, DAL = Dallas, DET = Detroit, LA = Los Angeles,} \\
\multicolumn{12}{l}{\footnotesize MIA = Miami, NYC = New York, PHI = Philadelphia, SF = San Francisco, SEA = Seattle, DC = Washington, DC.} \\
\multicolumn{12}{l}{\footnotesize All predictors standardized within city. Standard errors in parentheses. * p$<$0.05, ** p$<$0.01, *** p$<$0.001.} \\
\end{tabular}
}
\label{tab:regressions_scaled}
\end{table*}

\clearpage
\section{Place segregation of social infrastructures}
\label{SI:sec:social_infrastructures}
As described in the manuscript, we manually classify each POI into 3 categories, namely community places, social businesses, and parks or public outdoor spaces, to examine the relationship between experienced partisan segregation and visitation patterns to social infrastructures.
Table \ref{tab:social_infrastructures} reports the POI categories included in each group.

\begin{figure*}[ht!]
\centering
\includegraphics[width=1\textwidth]{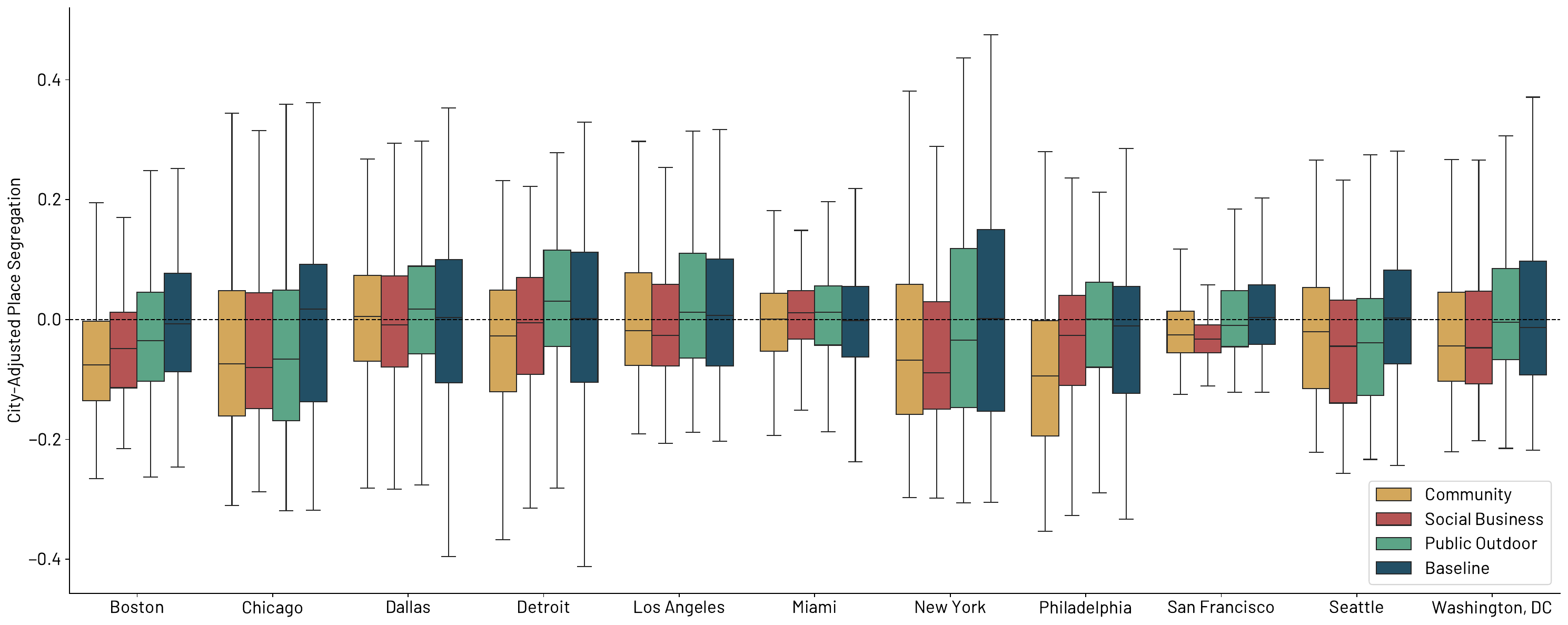} 
\caption{\textbf{Place segregation of social infrastructures across cities shows heterogeneity across cities.}}
\label{fig:SI_social}
\end{figure*}

Fig. \ref{fig:SI_social} shows the distributions of city-adjusted place partisan segregation for each group of social infrastructures, suggesting some variation across CBSAs.

Finally, Table \ref{tab:regr_social} reports the regressions' summaries of the analysis related to the relationship between visitation patterns of social infrastructures and experienced partisan segregation.

\begin{table}[h]
\centering

\begin{tabular}{@{}l | l@{}}
\hline

\parbox[t]{6cm}{\textbf{Social infrastructures group}} &
\parbox[t]{9cm}{\textbf{List of categories}} \\

\hline

\parbox[t]{6cm}{Community POIs} &
\parbox[t]{9cm}{
Community College, Community Center, Library, Meeting Room, City Hall, Town Hall, Science Museum, Art Museum, Museum, History Museum, Art Gallery, Arts, Art Studio, Arts \& Crafts, Performing Arts, Public Art, Street Art, Non-Profit, Cultural Center, Convention Center, University, Amphitheater, Capitol Building, Rec Center, Auditorium, Event Space, Administrative Building, Academic Building, Music School
} \\

\hline

\parbox[t]{6cm}{Social business POIs} &
\parbox[t]{9cm}{
Café, Bar, Beach Bar, Pet Café, Bookstore, College Bookstore, Used Bookstore, Sports Bar, Farmer's Market, Concert Hall, Theater, Music Festival, Aquarium, Opera House, Planetarium, Roof Deck, Festival, Zoo, Social Club, Arts \& Entertainment, Flea Market, Karaoke, Music Venue, Speakeasy, Coworking Space, Rock Club
} \\

\hline

\parbox[t]{6cm}{Public outdoor POIs} &
\parbox[t]{9cm}{
Park, Skate Park, Water Park, Bowling Green, Badminton Court, Waterfall, Cricket, Forest, National Park, Climbing Gym, Outdoor Event Space, Canal, Tennis, Botanical Garden, Indoor Play Area, Roller Rink, Mini Golf, Fountain, Outdoor Gym, Beach, Hockey, Hockey Field, Basketball, Pedestrian Street/Plaza, Rock Climbing, Volleyball Court, Sports Club, Basketball Court, Outdoors \& Recreation, Sculpture, Harbor / Marina, Stadium, Ski Area, Yoga Studio, Baseball, Football, River, Boxing Gym, Lake, Bowling Alley, Arcade, Skating Rink, Billiards, Golf Course, Soccer, Bathing Area, Soccer Field, Dance Studio, Historic Site, Garden, Gymnastics Gym, Pool, Athletics \& Sports, Tennis Court, Baseball Field, Gym, Gym / Fitness, Pilates Studio
} \\

\hline
\end{tabular}

\caption{Classification of points of interest into social infrastructure groups.}
\label{tab:social_infrastructures}

\end{table}

\clearpage

\begin{table*}[htbp]
\centering
\caption{Regression results of the relationship between patterns of use of social infrastructures and experienced partisan segregation across metropolitan areas - Number of POIs within a 15-minute walking distance.}
\resizebox{\textwidth}{!}{%
\begin{tabular}{lccccccccccc}
\toprule
& \multicolumn{11}{c}{\textbf{Metropolitan Areas}} \\
\cmidrule(lr){2-12}
& BOS & CHI & DAL & DET & LA & MIA & NYC & PHI & SF & SEA & DC \\
\midrule

$S_{i}^{res}$
& 0.359*** & 0.302*** & 0.305*** & 0.497*** & 0.479*** & 0.209*** & 0.344*** & 0.296*** & 0.443*** & 0.418*** & 0.529*** \\
& (0.004) & (0.002) & (0.002) & (0.005) & (0.002) & (0.002) & (0.003) & (0.004) & (0.003) & (0.003) & (0.003) \\

Log POIs: Community
& 0.002 & -0.009*** & 0.010*** & 0.013*** & -0.006*** & -0.004*** & -0.013*** & -0.044*** & -0.002* & -0.012*** & -0.006*** \\
& (0.002) & (0.001) & (0.001) & (0.003) & (0.001) & (0.001) & (0.001) & (0.002) & (0.001) & (0.002) & (0.001) \\

Log POIs: Social Business
& 0.007*** & 0.004*** & -0.004* & -0.005 & -0.003*** & -0.001 & 0.029*** & 0.009*** & -0.003*** & 0.006*** & -0.002 \\
& (0.002) & (0.001) & (0.001) & (0.003) & (0.001) & (0.001) & (0.001) & (0.002) & (0.001) & (0.002) & (0.001) \\

Log POIs: Public Outdoor
& -0.029*** & -0.059*** & -0.025*** & -0.005 & -0.005*** & -0.007*** & -0.056*** & -0.009*** & -0.001 & 0.010*** & 0.008*** \\
& (0.002) & (0.001) & (0.001) & (0.002) & (0.001) & (0.001) & (0.001) & (0.002) & (0.001) & (0.002) & (0.001) \\

Time Spent: Community
& -0.188*** & -0.090*** & 0.035*** & -0.059*** & -0.014** & -0.002 & -0.134*** & -0.151*** & -0.073*** & -0.107*** & -0.072*** \\
& (0.008) & (0.006) & (0.007) & (0.012) & (0.005) & (0.006) & (0.010) & (0.010) & (0.006) & (0.010) & (0.009) \\

Time Spent: Social Business
& -0.133*** & -0.049*** & -0.028*** & -0.044*** & -0.017*** & 0.038*** & -0.123*** & 0.008 & -0.131*** & -0.168*** & -0.063*** \\
& (0.009) & (0.005) & (0.006) & (0.009) & (0.005) & (0.004) & (0.008) & (0.008) & (0.005) & (0.009) & (0.007) \\

Time Spent: Public Outdoor
& -0.060*** & -0.054*** & -0.021*** & 0.019* & 0.020*** & 0.020*** & -0.016** & -0.024** & -0.024*** & -0.088*** & -0.036*** \\
& (0.006) & (0.004) & (0.004) & (0.008) & (0.003) & (0.003) & (0.006) & (0.008) & (0.004) & (0.006) & (0.005) \\

\% African Americans
& -0.073*** & -0.102*** & -0.185*** & 0.087*** & 0.004 & -0.057*** & -0.058*** & -0.011* & 0.099*** & 0.028** & -0.096*** \\
& (0.006) & (0.003) & (0.004) & (0.007) & (0.003) & (0.003) & (0.004) & (0.005) & (0.005) & (0.009) & (0.004) \\

\% Hispanic
& 0.008 & -0.103*** & -0.202*** & 0.052*** & 0.035*** & 0.000 & -0.081*** & 0.018* & 0.058*** & -0.038*** & -0.085*** \\
& (0.007) & (0.003) & (0.004) & (0.012) & (0.002) & (0.001) & (0.005) & (0.008) & (0.003) & (0.007) & (0.006) \\

Log Income
& -0.024*** & -0.007*** & -0.023*** & -0.033*** & 0.010*** & -0.002* & 0.012*** & 0.002 & 0.000 & -0.019*** & -0.063*** \\
& (0.002) & (0.001) & (0.002) & (0.003) & (0.001) & (0.001) & (0.002) & (0.003) & (0.001) & (0.002) & (0.002) \\

\% Labor Force
& 0.011 & 0.174*** & 0.092*** & 0.112*** & 0.128*** & -0.012*** & 0.066*** & 0.074*** & 0.045*** & 0.011 & 0.154*** \\
& (0.007) & (0.004) & (0.004) & (0.009) & (0.004) & (0.003) & (0.006) & (0.006) & (0.004) & (0.006) & (0.005) \\

\% Graduates
& -0.031*** & -0.052*** & -0.091*** & 0.036*** & -0.020*** & 0.031*** & -0.104*** & 0.042*** & 0.029*** & -0.113*** & -0.052*** \\
& (0.005) & (0.003) & (0.003) & (0.006) & (0.003) & (0.003) & (0.005) & (0.006) & (0.003) & (0.005) & (0.005) \\

Poverty Ratio
& 0.029** & 0.056*** & 0.044*** & -0.031** & 0.095*** & 0.017*** & -0.054*** & 0.005 & 0.012 & 0.013 & -0.026** \\
& (0.010) & (0.005) & (0.005) & (0.010) & (0.004) & (0.004) & (0.009) & (0.010) & (0.006) & (0.009) & (0.008) \\

\midrule
Num.\ Obs.
& 40,124 & 190,213 & 169,326 & 72,857 & 176,903 & 127,343 & 158,261 & 76,921 & 56,023 & 50,184 & 101,284 \\

$R^2$
& 0.585 & 0.627 & 0.577 & 0.624 & 0.612 & 0.420 & 0.604 & 0.537 & 0.601 & 0.670 & 0.705 \\

\bottomrule
\multicolumn{12}{l}{\footnotesize \textit{Note:} BOS = Boston, CHI = Chicago, DAL = Dallas, DET = Detroit, LA = Los Angeles,} \\
\multicolumn{12}{l}{\footnotesize MIA = Miami, NYC = New York, PHI = Philadelphia, SF = San Francisco, SEA = Seattle, DC = Washington, DC.} \\
\multicolumn{12}{l}{\footnotesize Standard errors in parentheses. * p$<$0.05, ** p$<$0.01, *** p$<$0.001.} \\
\end{tabular}
}
\label{tab:regr_social}
\end{table*}

\clearpage
\begin{table*}[htbp]
\centering
\caption{Regression results of the relationship between patterns of use of social infrastructures and experienced partisan segregation across metropolitan areas - Number of POIs within a 30-minute walking distance.}
\resizebox{\textwidth}{!}{%
\begin{tabular}{lccccccccccc}
\toprule
& \multicolumn{11}{c}{\textbf{Metropolitan Areas}} \\
\cmidrule(lr){2-12}
& BOS & CHI & DAL & DET & LA & MIA & NYC & PHI & SF & SEA & DC \\
\midrule

$S_{i}^{res}$
& 0.328*** & 0.259*** & 0.302*** & 0.489*** & 0.475*** & 0.210*** & 0.300*** & 0.257*** & 0.434*** & 0.422*** & 0.534*** \\
& (0.004) & (0.002) & (0.002) & (0.005) & (0.002) & (0.002) & (0.003) & (0.004) & (0.003) & (0.003) & (0.003) \\

Log POIs: Community
& 0.008*** & -0.012*** & 0.004*** & 0.014*** & -0.013*** & -0.003*** & -0.011*** & -0.045*** & -0.003*** & -0.004** & -0.011*** \\
& (0.001) & (0.001) & (0.001) & (0.002) & (0.001) & (0.001) & (0.001) & (0.001) & (0.001) & (0.001) & (0.001) \\

Log POIs: Social Business
& 0.007*** & 0.002* & 0.006*** & -0.013*** & 0.000 & 0.001* & 0.022*** & 0.006*** & -0.004*** & 0.008*** & 0.004*** \\
& (0.001) & (0.001) & (0.001) & (0.002) & (0.001) & (0.001) & (0.001) & (0.001) & (0.001) & (0.001) & (0.001) \\

Log POIs: Public Outdoor
& -0.038*** & -0.045*** & -0.024*** & -0.013*** & -0.000 & -0.006*** & -0.059*** & -0.008*** & -0.000 & 0.002* & 0.010*** \\
& (0.001) & (0.001) & (0.001) & (0.001) & (0.000) & (0.001) & (0.001) & (0.001) & (0.001) & (0.001) & (0.001) \\

Time Spent: Community
& -0.179*** & -0.079*** & 0.036*** & -0.059*** & -0.010* & -0.001 & -0.110*** & -0.129*** & -0.071*** & -0.110*** & -0.069*** \\
& (0.008) & (0.006) & (0.007) & (0.012) & (0.005) & (0.006) & (0.010) & (0.010) & (0.006) & (0.010) & (0.009) \\

Time Spent: Social Business
& -0.124*** & -0.040*** & -0.026*** & -0.039*** & -0.016*** & 0.038*** & -0.097*** & 0.024** & -0.125*** & -0.169*** & -0.066*** \\
& (0.009) & (0.005) & (0.006) & (0.009) & (0.005) & (0.004) & (0.008) & (0.007) & (0.005) & (0.009) & (0.007) \\

Time Spent: Public Outdoor
& -0.050*** & -0.046*** & -0.018*** & 0.022** & 0.020*** & 0.020*** & -0.006 & -0.017* & -0.023*** & -0.088*** & -0.037*** \\
& (0.005) & (0.004) & (0.004) & (0.008) & (0.003) & (0.003) & (0.006) & (0.008) & (0.004) & (0.006) & (0.005) \\

\% African Americans
& -0.090*** & -0.132*** & -0.190*** & 0.074*** & 0.001 & -0.057*** & -0.089*** & -0.024*** & 0.093*** & 0.029** & -0.087*** \\
& (0.006) & (0.003) & (0.004) & (0.007) & (0.003) & (0.003) & (0.004) & (0.005) & (0.005) & (0.009) & (0.004) \\

\% Hispanic
& 0.002 & -0.113*** & -0.200*** & 0.046*** & 0.036*** & 0.001 & -0.103*** & -0.011 & 0.053*** & -0.033*** & -0.084*** \\
& (0.006) & (0.003) & (0.004) & (0.012) & (0.002) & (0.001) & (0.004) & (0.008) & (0.003) & (0.007) & (0.006) \\

Log Income
& -0.027*** & -0.011*** & -0.025*** & -0.039*** & 0.008*** & -0.003* & -0.001 & -0.013*** & -0.002 & -0.018*** & -0.060*** \\
& (0.002) & (0.001) & (0.002) & (0.003) & (0.001) & (0.001) & (0.002) & (0.003) & (0.001) & (0.002) & (0.002) \\

\% Labor Force
& 0.025*** & 0.167*** & 0.096*** & 0.124*** & 0.129*** & -0.010** & 0.074*** & 0.065*** & 0.052*** & 0.010 & 0.144*** \\
& (0.007) & (0.004) & (0.004) & (0.009) & (0.004) & (0.003) & (0.005) & (0.006) & (0.004) & (0.006) & (0.005) \\

\% Graduates
& -0.035*** & -0.038*** & -0.078*** & 0.043*** & -0.015*** & 0.031*** & -0.090*** & 0.057*** & 0.034*** & -0.113*** & -0.059*** \\
& (0.005) & (0.003) & (0.003) & (0.006) & (0.003) & (0.003) & (0.005) & (0.005) & (0.003) & (0.005) & (0.005) \\

Poverty Ratio
& 0.033*** & 0.057*** & 0.040*** & -0.041*** & 0.098*** & 0.014*** & -0.034*** & 0.022* & 0.018** & 0.008 & -0.028*** \\
& (0.010) & (0.005) & (0.005) & (0.010) & (0.004) & (0.004) & (0.008) & (0.009) & (0.006) & (0.009) & (0.008) \\

\midrule
Num.\ Obs.
& 40,124 & 190,213 & 169,326 & 72,857 & 176,903 & 127,343 & 158,261 & 76,921 & 56,023 & 50,184 & 101,284 \\

$R^2$
& 0.597 & 0.642 & 0.579 & 0.626 & 0.614 & 0.421 & 0.625 & 0.558 & 0.603 & 0.670 & 0.706 \\

\bottomrule
\multicolumn{12}{l}{\footnotesize \textit{Note:} BOS = Boston, CHI = Chicago, DAL = Dallas, DET = Detroit, LA = Los Angeles,} \\
\multicolumn{12}{l}{\footnotesize MIA = Miami, NYC = New York, PHI = Philadelphia, SF = San Francisco, SEA = Seattle, DC = Washington, DC.} \\
\multicolumn{12}{l}{\footnotesize Standard errors in parentheses. * p$<$0.05, ** p$<$0.01, *** p$<$0.001.} \\
\end{tabular}
}
\label{tab:regr_social_30min}
\end{table*}

% \bibliographystyle{plain}
% \bibliography{sample}